\pdfoutput=1
\documentclass[11pt]{article}

\usepackage[]{EMNLP2023}
\usepackage{times}
\usepackage{latexsym}

\usepackage[T1]{fontenc}
\usepackage[utf8]{inputenc}

\usepackage{microtype}

\usepackage{inconsolata}

\usepackage{multirow}
\usepackage{graphicx}
\usepackage{multirow}
\usepackage{booktabs}
\usepackage{tablefootnote}
\usepackage{subcaption}
\usepackage{cleveref}

\title{
Large Language Models as Test Case Generators: Performance Evaluation and Enhancement 
}

\author{Kefan Li \\
  Beihang University \\
  \\\And
  Yuan Yuan \\
  Beihang University \\
  \\}

\begin{document}
\maketitle
\begin{abstract}

Code generation with Large Language Models (LLMs) has been extensively studied and achieved remarkable progress.
As a complementary aspect to code generation, test case generation is of crucial importance 
in ensuring the quality and reliability of code. However, using LLMs as test case generators has been much less explored. 
Current research along this line primarily focuses on enhancing code generation with assistance from test cases 
generated by LLMs, while the performance of LLMs in test case generation alone has not been comprehensively examined. 
To bridge this gap, we conduct extensive experiments to study how well LLMs can generate high-quality test cases. 
We find that as the problem difficulty increases, state-of-the-art LLMs struggle to generate correct test cases, largely 
due to their inherent limitations in computation and reasoning. To mitigate this issue, we further propose
a multi-agent framework called \emph{TestChain} that decouples the generation of test inputs and test outputs.
Notably, TestChain uses a ReAct format conversation chain for LLMs to interact with a Python interpreter in order to
provide more accurate test outputs. Our results indicate that TestChain outperforms the baseline by a large margin. 
Particularly, in terms of the accuracy of test cases, 
TestChain using GPT-4 as the backbone achieves a 13.84\% improvement over the baseline on the LeetCode-hard dataset.
\end{abstract}

\section{Introduction}
Large Language Models (LLMs) have shown significant capabilities in code generation, paving the way 
for transformative changes in software development. Among these LLMs, there are those specifically designed for coding, such as Codex \cite{chen2021evaluating},
CodeGen \cite{nijkamp2022codegen} and CodeLlama \cite{roziere2023code}, as well as general-purpose models like GPT-3.5 and GPT-4 \cite{achiam2023gpt}.

\begin{figure}[!ht]
\begin{center}
\centering
\includegraphics[width=\linewidth]{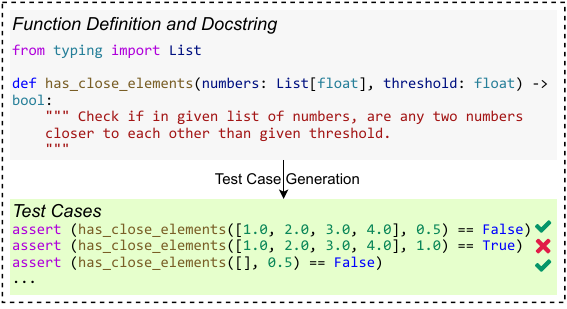}
\caption{
Formulation of test case generation in this paper, where an assertion encapsulates a test case.
}
\label{fig:testcase}
\end{center}
\end{figure}

Test case generation usually refers to the automatic process of creating test cases. 
It is essential for ensuring the quality and reliability of code, serving as a complementary aspect to code generation.
Note that the concept of test cases spans various levels (unit testing, system testing, etc.) and scopes (functional testing, security testing, etc.) within the software testing process.
Following the recent related studies \cite{chen2022codet,huang2023codecot,huang2023agentcoder,shinn2023reflexion}, 
we consider function-level unit test cases in this paper, where \emph{a test case} refers to a pair of input and expected output for the function defined in the given context. 
Figure~\ref{fig:testcase} describes the formulation of test case generation.
%where an assertion encapsulates a test case. 

Because of breakthroughs in code generation achieved by LLMs, it is intuitive to apply these models to test case generation as well.
This type of research has only recently emerged as a field of interest.
CodeT \cite{chen2022codet} uses the test cases generated by LLMs
to select a single solution from multiple code samples. 
In CodeCoT \cite{huang2023codecot}, Reflexion \cite{shinn2023reflexion} and AgentCoder \cite{huang2023agentcoder}, 
the execution of programs on test cases generated by LLMs provides feedback information for self-correction.

However, most of current research in this direction primarily focuses on utilizing test cases to benefit code generation, with little attention to the quality of test cases generated by LLMs.
If the test cases generated by LLMs contain many errors, the effectiveness of these approaches will be significantly diminished.
A great challenge in test case generation is how to accurately map each test input to its corresponding output, often requiring precise mathematical calculations and complex logical reasoning.
Nonetheless, existing studies \cite{frieder2023mathematical,liu2023evaluating} have shown that even the most capable LLMs are not proficient in this regard.
Thus, it naturally leads us to pose the following question: \emph{Can LLMs really generate high-quality test cases?}

In this paper, we first conduct extensive experiments to explore this question, where the quality of test cases is evaluated by comprehensive metrics including accuracy, line coverage and our proposed
new metric called \emph{Code-with-Bugs} concerning the strength of tests. 
We examine four notable LLMs comprising both open and closed-source frameworks. Specifically, we include two open-source models: StarChat \cite{li2023starcoder} and CodeLlama \cite{roziere2023code}, 
alongside two close-source models: GPT-3.5 and GPT-4.
We choose a relatively easy dataset HumanEval \cite{chen2021evaluating} and a more challenging dataset LeetCode-hard \cite{shinn2023reflexion} for evaluation.
Experiments show that StarChat and CodeLlama struggle to generate correct test cases, while GPT-3.5 and GPT-4 perform well on HumanEval but tend to produce many incorrect test cases on LeetCode-hard.

Based on the insights from the empirical evaluation, we further propose a multi-agent framework called \emph{TestChain}, 
in order to enhance the performance of LLMs in test case generation. 
In TestChain, one agent called \emph{Designer agent} is responsible for generating diverse test inputs, while another agent called \emph{Calculator agent} is responsible for correctly mapping the inputs to outputs.
Notably, Calculator agent uses a ReAct \cite{yao2022react} format conversation chain interaction with a Python interpreter for executing code, 
which can significantly reduce the complexity and inaccuracy in the input-output mapping. 
Experiments demonstrate that our approach significantly outperforms the baseline in terms of all metrics.  
In summary, the main contributions of this paper are twofold:
\begin{itemize}
\item We conduct a comprehensive evaluation of LLMs in test case generation. We find that the performance of LLMs 
drops sharply in handling harder problems, where the generated test cases may contain many errors. 

\item We propose \emph{TestChain}, which significantly enhances the performance of LLMs in generating test cases.
TestChain achieves an accuracy of \textbf{71.79\%} with GPT-4 on the LeetCode-hard dataset, while the Test Agent (1-shot) baseline method only gets an accuracy of \textbf{57.95\%}.
\end{itemize}

\section{Related Work}
\textbf{LLMs for Test Case Generation.}
Recently, there has been a growing trend of utilizing LLMs for test case generation. 
\citet{tufano2020unit} and \citet{li2022competition} utilize pre-trained language models fine-tuned on labeled data to generate test cases. CodeT \cite{chen2022codet} directly generates test cases with the powerful model Codex using zero-shot prompts. CodeCoT \cite{huang2023codecot} employs a Chain of Thought prompt to generate both solutions and test cases for programming tasks. Reflexion and AgentCoder employ LLMs to generate test cases and execute them on programs to produce feedback information for self-correction. 

Automatically generating test cases with LLMs can also be used in real-world software development. \citet{zhang2023well} utilizes GPT-4 to generate security tests to find vulnerabilities. \citet{schafer2023empirical} employs LLMs to assist in automatic software testing by generating unit tests. \citet{kang2023large} utilizes LLMs to generate test cases to reproduce general bugs in software.

\textbf{Multi-Agent Structure with LLMs.}
Utilizing multiple agents driven by LLMs to collaborate in solving tasks has emerged as a popular approach. Self-Collaboration \cite{dong2023self}  establishes a team comprising analyst, coder, and tester roles for software development. INTERVENOR \cite{wang2023intervenor} employs a learner agent for generating programs and a teacher agent for suggesting repairs. 
AgentCoder \cite{huang2023agentcoder} introduces a programmer agent, a test designer agent, and a test executor agent for code generation and refinement, with the programmer agent and the test designer agent powered by LLMs. AutoGen \cite{wu2023autogen} provides a straightforward method for creating multi-agent applications with LLMs.

\textbf{LLMs Interaction with Tools.}
By defining specific conversation formats, the generations of LLMs can be parsed into tool invocations, enabling interaction between LLMs and external tools.
ReAct \cite{yao2022react} employs a chain in ``Thought/Action/Observation'' format, allowing LLMs to provide tool types and return the execution results of these tools to LLMs.
Recently, an open-source project LangChain\footnote{https://python.langchain.com} allows the creation of LLM applications with ReAct or custom-defined conversation chains.
FunSearch \cite{FunSearch2023} utilizes Google's PaLM 2 \cite{anil2023palm} and customized evaluators to discover new solutions to the cap set problem.
HuggingGPT \cite{shen2023hugginggpt} employs GPT-4 and numerous expert models to solve AI tasks in a collaborative way.
WebGPT \cite{nakano2021webgpt} allows LLMs to use a browser to assist in answering questions.
AutoGen can create a Python interpreter environment locally or via docker and interact with LLM agents.
Research based on AutoGen can create conversations to conduct hyperparameter Optimization \cite{wang2023EcoOptiGen} and solve math problems \cite{wu2023empirical}.

\begin{table*}[ht]
\renewcommand{\arraystretch}{1.0}
\centering
\footnotesize
\tabcolsep =6pt
\begin{tabular}{llcccccc}
\toprule
\multirow{2}{*}{\textbf{Model}} & \multirow{2}{*}{\textbf{Method}} & \multicolumn{3}{c}{\textbf{HumanEval-no-exp}} && \multicolumn{2}{c}{\textbf{LeetCode-no-exp}} \\
\cmidrule{3-5}
\cmidrule{7-8}
& & \textbf{Accuracy} (\%) & \textbf{Line Cov} (\%) & \textbf{CwB} (\%) && \textbf{Accuracy} (\%) & \textbf{Line Cov} (\%)\\
\midrule
\multirow{2}{*}{StarChat}
& Test Agent (0-shot) & 37.44 & 51.86 & 45.61 && 9.74 & 31.10 \\
& Test Agent (1-shot)  & 41.10 & 52.27 & 45.46 && 12.31 & 23.61 \\
\midrule
\multirow{2}{*}{CodeLlama}
& Test Agent (0-shot) & 51.83 & 63.88 & 54.85 && 23.08 & 58.61 \\
& Test Agent (1-shot)  & 67.07 & 71.71 & 63.23 && 28.72 & 55.95 \\
\midrule
\multirow{2}{*}{GPT-3.5}
& Test Agent (0-shot) & 65.98 & 71.60 & 69.60 && 26.67 & 63.68 \\
& Test Agent (1-shot)  & 74.02 & 74.69 & 74.15 && 38.97 & 73.66 \\
\midrule
\multirow{2}{*}{GPT-4}
& Test Agent (0-shot) & 84.02 & 77.54 & 82.53 && 58.97 & 88.83 \\
& Test Agent (1-shot) & 84.63 & 77.04 & 83.11 && 57.95 & 88.47 \\
\bottomrule
\end{tabular}
\caption{
Evaluation results of the Tester Agent with 0-shot and 1-shot prompt.
}
\label{tab:table1}
\end{table*}

\section{Can LLMs Really Generate High-Quality Test Cases?}

To thoroughly investigate the capabilities of LLMs in generating test cases, we select four LLMs and two datasets for experimentation.
We employ three metrics to evaluate the quality of test cases from different aspects, including correctness and strength.

\subsection{Experimental Setup}
\label{sec:can-experimental-setup}

\textbf{Models.}
The four typical LLMs include two open-source models: \emph{StarChat-beta} and \emph{CodeLlama-13b-Instruct-hf}, and two closed-source models: \emph{GPT-3.5-turbo-1106} and \emph{GPT-4-turbo-preview}.
In the following, we refer to the four models by \emph{StarChat}, \emph{CodeLlama}, \emph{GPT-3.5} and \emph{GPT-4} in short.

\textbf{Datasets.}
We choose a popular Python programming dataset HumanEval that contains a canonical solution for each question.
Moreover, we choose a more challenging dataset LeetCode-hard that includes $39$ difficult Python programming tasks.
For some questions in the original LeetCode-hard, the canonical solution is either missing or faulty.
So we manually collect correct solutions for them and verify these solutions via the LeetCode website.

We have also noticed that LLMs may plagiarize the test case examples included in the question prompt, so we remove all the examples from the prompts of the two datasets.
Examples of prompts with and without examples can be found in Appendix~\ref{sec:dataset-examples}.
In the following, we refer to the two datasets without examples by \emph{HumanEval-no-exp} and \emph{LeetCode-no-exp}.

\textbf{Generation Setup.}
For using LLMs as test case generators, we design a \emph{system prompt}
that sets the LLM as a Python tester and instructs it
to generate \emph{basic}, \emph{edge}, and \emph{large scale} test cases \cite{huang2023agentcoder} for a given context.
We consider two paradigms for generation: \emph{Test Agent} with \emph{0-shot} and \emph{1-shot} prompt.
For 1-shot generation, examples illustrating the test case generation are
prefixed to the user prompt. 
Please refer to Figures~\ref{fig:system-prompt}--\ref{fig:few-shot-prompt}
in Appendix~\ref{sec:prompt-examples} for the details of the prompts used in 
our experiments. 

For each question, we remove duplicate and syntactically erroneous test cases. If more than five remain, we retain the first five for evaluation; otherwise, we keep all.
The HumanEval-no-exp dataset comprises 164 questions, with a maximum of 820 test cases. The LeetCode-no-exp dataset includes 39 questions, with a maximum of 195 test cases.

\textbf{Metrics.}
A test case is considered to be correct if it can be passed by the canonical solution. 
The accuracy measures the percentage of correct test cases among all those generated.
To comprehensively assess the quality of test cases for a question, line coverage (Line Cov) is used as another metric, measuring the proportion of lines accessed 
when the test cases are executed on the canonical solution. For comparison, we calculate the average line coverage on the full dataset.

Line coverage alone is recognized as insufficient for evaluating the strength of test cases \cite{andrews2006using}.
We also introduce a new metric called \emph{Code-with-Bugs} pass rate (CwB in short).
To compute CwB for a set of test cases, we first randomly 
generate a number of faulty programs (20 in our experiments) for the question.
Then, CwB is the proportion of programs that fail this test set.
We use the average CwB on the full dataset for comparison. 
Note that to conduct a fair comparison, we utilize another model, 
CodeGen-Mono 6B \cite{nijkamp2022codegen}, to generate faulty programs. 
Additionally, CwB is only applied to HumanEval in our experiments, as the canonical test cases for verifying correctness are not available for LeetCode-hard.

To further investigate why LLMs generate incorrect test cases, we analyze error types,
categorizing them into \emph{Assertion Error}, \emph{Runtime Error} and \emph{Timeout Error}:
\begin{itemize}
\item \textbf{Assertion Error}: Indicates a mismatch between the execution output of the target function and the expected output.
\item \textbf{Runtime Error}: Indicates an internal error during function execution, often due to non-compliant input.
\item \textbf{Timeout Error}: Indicates that the execution of the target function exceeded the allotted time limit (1 second in our experiments).
\end{itemize}
We will provide statistics on the occurrence of each error type.

\textbf{Parameters.}
For all the models, we use a temperature of $T=0.2$ and $top\_p$ is set to $0.95$.
The $max\_new\_tokens$ for HuggingFace models and the $max\_tokens$ for OpenAI models are set to $1024$.
For GPT-3.5 and GPT-4, we call them via the official OpenAI APIs \footnote{https://platform.openai.com}.
For StarChat and CodeLlama, we use HuggingFace transformers \cite{wolf2019huggingface} for inference.
We conduct experiments on two V100 GPUs with 32GB of memory each.

\subsection{Evaluation Results}
\label{sec:can-evaluation-results}

\begin{figure}[!ht]
\begin{center}
%\vskip 0.2in
\centering
\includegraphics[width=\linewidth]{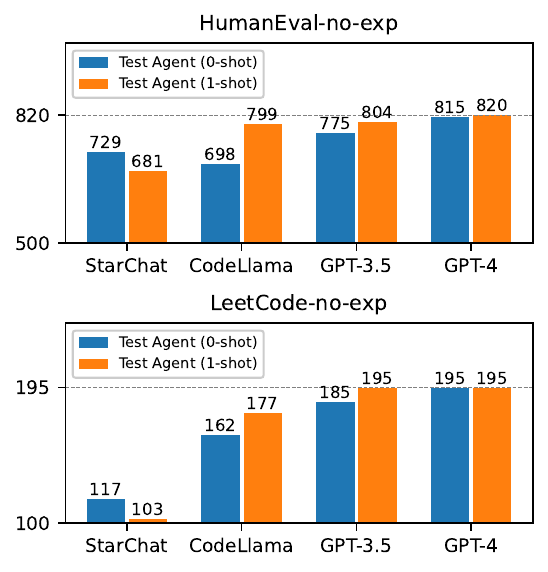}
\caption{
Statistics on the number of test cases successfully generated.
The gray line indicates the maximum number of test cases allowed by the dataset.
}
\label{fig:generated-1}
\end{center}
%\vskip -0.2in
\end{figure}

Figure~\ref{fig:generated-1} shows the number of test cases each model and method successfully generated. It reveals that, for the relatively easy HumanEval-no-exp dataset, models are capable of producing an ample number of test cases. Conversely, for the harder LeetCode-no-exp dataset,  less advanced models like StarChat face difficulties in generating a sufficient quantity of test cases.
This also implies that the most powerful models like GPT-3.5 and GPT-4 can capture the syntax of the test cases very well. 

\begin{figure*}[ht]
\begin{center}
\begin{subfigure}{2\columnwidth}
\vskip 0.2in
\centering
\includegraphics[width=\linewidth]{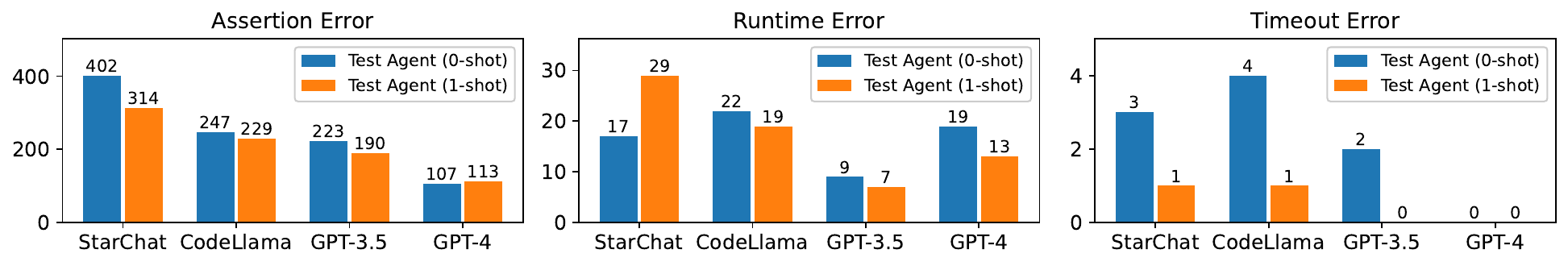}
\caption{
Statistics on the HumanEval-no-exp dataset.
}
\vskip -0.2in
\end{subfigure}
\begin{subfigure}{2\columnwidth}
\vskip 0.2in
\centering
\includegraphics[width=\linewidth]{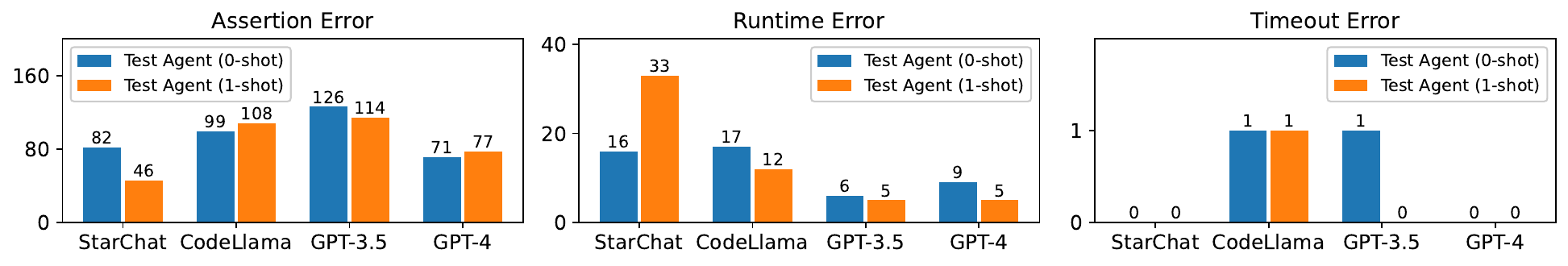}
\caption{
Statistics on the LeetCode-no-exp dataset.
}
\vskip -0.2in
\end{subfigure}
\caption{
Statistics on the number of incorect test cases for each type of error.
}
\label{fig:counts-1}
\end{center}
\end{figure*}

Table~\ref{tab:table1} presents the evaluation results across all models and datasets. 
The results reveal that the Test Agent (1-shot) method generally outperforms the zero-shot method in terms of accuracy. 
This is particularly clear for CodeLlama on HumanEval-no-exp (a 15.24\% improvement) and GPT-3.5 on LeetCode-no-exp (a 12.3\% improvement).
While in terms of test strength, the situation varies significantly between different models.
For example, 1-shot is of great help to GPT-3.5 in achieving an improved test strength, while it
is unhelpful or even harmful to StarChat in this respect. 
It is worth noting that, for GPT-4, there is little difference between 0-shot and 1-shot in all metrics. 
This suggests that for the particularly powerful models like GPT-4, the illustrating examples may not be necessary. 

In Table~\ref{tab:table1}, we can also observe a sharp decline in accuracy for all the generation methods from 
HumanEval-no-exp to LeetCode-no-exp. 
Even the accuracy of GPT-4 decreases by about 25\%,  
reaching only about 58\% on LeetCode-hard. 
This underscores the need for further advancements in test case generation, particularly for hard questions.
In terms of line coverage, there has been a notable decrease observed in StarChat and CodeLlama, 
whereas GPT-3.5 and GPT-4 have not experienced a similar decline. 
This indicates that the overall quality of test cases 
generated by the most advanced models like GPT-4 is bottlenecked by accuracy rather than coverage.

Figure~\ref{fig:counts-1} shows the number of incorrect test cases for each type of error.
It can be seen that Assertion Error is the most prevalent type of error, while the other types of errors constitute only a small fraction.
This suggests that the trickiest issue for generating correct test cases lies
in accurately computing the corresponding test outputs given the test inputs.

%In short, LLMs are capable of generating a considerable number of correct test cases for easy questions, such as those in HumanEval.
%However, for questions concerning complex computation and reasoning, such as those in LeetCode-hard, 
%even the most advanced models like GPT-4 struggle to generate correct test cases. 
%This motivates us to use the external computational tools to address the computation and reasoning problem, 
%so as to enhance the performance of LLMs in test case generation. 

In summary, LLMs are capable of generating a considerable number of correct test cases for relatively easy questions, such as those in HumanEval.
However, for harder questions such as those in LeetCode-hard, all of them experience a sharp decline in accuracy and 
struggle to generate correct test cases. 
It becomes evident that the pivotal factor in elevating the quality of generated test cases is the enhancement of their correctness, 
as the overall quality is primarily constrained by accuracy for the advanced LLMs like GPT-4. 
The correctness of test cases essentially hinges on the accurate mapping of test inputs to their outputs, a process that may involve complex calculations and reasoning, areas where LLMs do not excel.
This motivates us to decompose this mapping problem into a number of steps to reduce the complexity, and to utilize external tools to enhance the precision of computation and reasoning.

\begin{figure*}[!ht]
%\vskip 0.2in
\begin{center}
\centerline{\includegraphics[width=2\columnwidth]{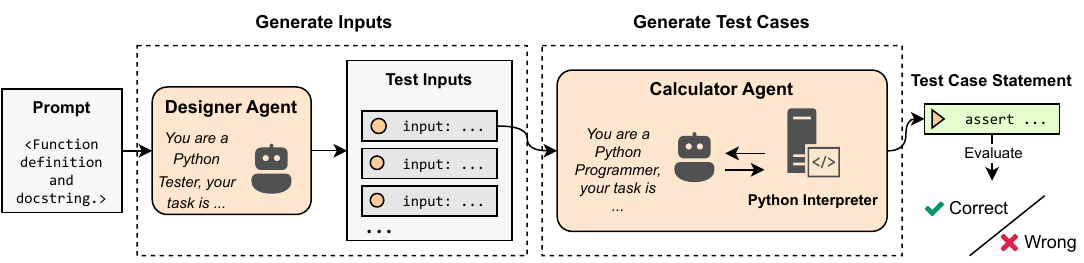}}
\caption{
Illustration of the TestChain framework.
}
\label{fig:agents}
\end{center}
%\vskip -0.2in
\end{figure*}

\section{Methodology}
\subsection{TestChain Framework}
\label{sec:testchain-framework}

Based on the above insights, we propose the \emph{TestChain} framework that allows LLMs to generate test cases with the assistance of a Python interpreter.
TestChain uses the divide-and-conquer idea to decompose the test case generation into two sequential sub-tasks: test input generation and test output generation. 
The two subtasks are handled by two agents called \emph{Designer agent} and \emph{Calculator agent}, respectively. 
Because the test input-output mapping may involve complex computation 
and reasoning, TestChain divides the mapping process into a number of small steps using a conservation chain similar to ReAct \cite{yao2022react}.
In each step, the Python code snippet will be written by the LLM and then be executed by the Python interpreter, in order to
achieve the desired goal. Figure~\ref{fig:agents} illustrates our TestChain framework.

It should be noted that in the generation process, Designer agent will be called only once to generate diverse test inputs, but Calculator agent will be called for each test input.

\begin{figure*}[!ht]
\vskip 0.2in
\begin{center}
\centerline{\includegraphics[width=2\columnwidth]{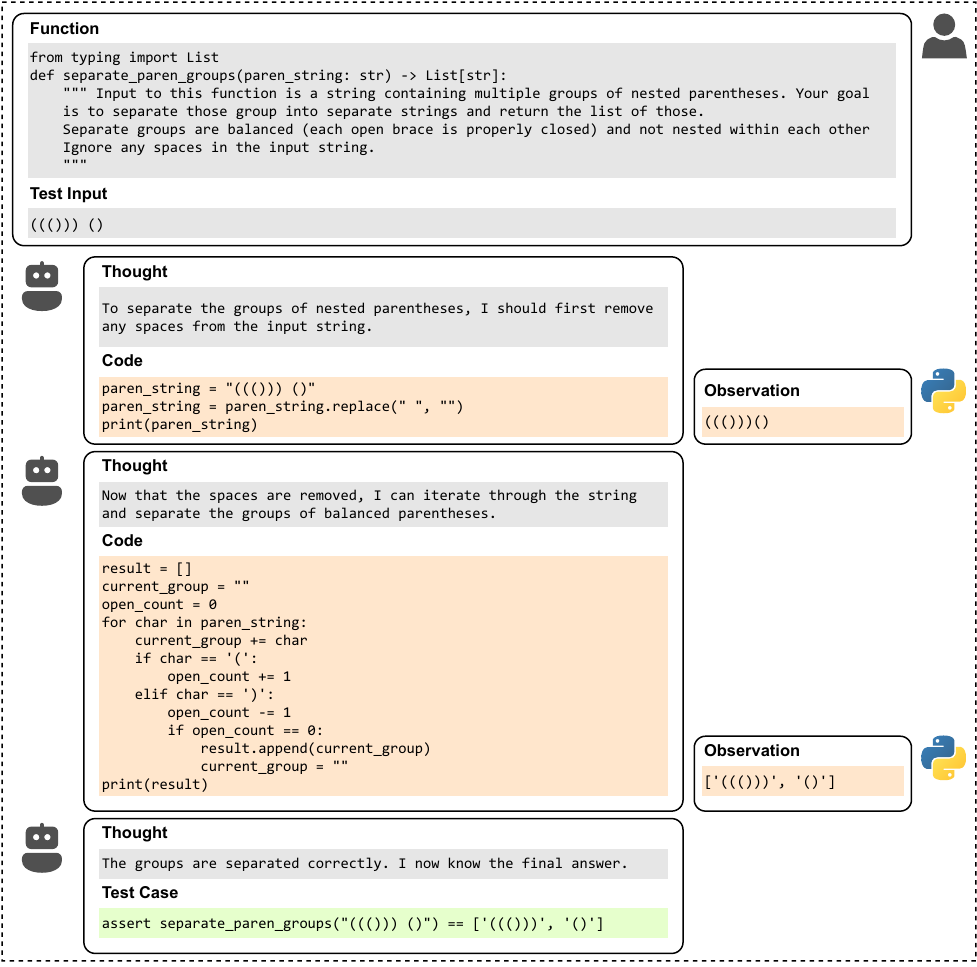}}
\caption{
Example of the conversation process produced by Calculator agent.
}
\label{fig:conversation}
\end{center}
\vskip -0.2in
\end{figure*}

\subsection{Designer Agent}
\label{sec:designer-agent}
Designer agent employs the LLM for test input generation.
In this agent, the system prompt describes that it will generate basic and edge test inputs.
The 1-shot prompt provides an example of the two types of test inputs 
in addition to the function definition and docstring. 
Please refer to Figures~\ref{fig:designer-agent-system-prompt}--\ref{fig:designer-agent-few-shot-prompt}
in Appendix~\ref{sec:prompt-examples} for the details of the prompts used by Designer agent. 
After generation, the duplicated test inputs will be omitted and the retained ones will be the input to Calculator agent.

Note that Designer agent does not need to concern itself with test outputs; 
it only needs to generate test inputs rather than complete test cases. 
With this approach, the LLM can concentrate on producing diverse test inputs without the necessity of computing the corresponding test outputs, thereby minimizing the likelihood of errors.

%In our setup, a line starting with a ``input:'' indicates a test input.
%The prompts used in the designer agent are similar to the few-shot method but have the following differences: (1) The test case generation definition in the system prompt describes that it will generate basic, edge, and large scale test inputs. (2) The few-shot prompt of the designer agent put the test inputs into a text block but not a Python code block.
%Prompts for the designer agent are shown in Appendix~\ref{sec:prompt-examples}.
%Therefore, the designer agent does not need to predict the output of test cases but only needs to generate test inputs instead of complete test cases.
%Through this approach, the LLMs can focus on generating diverse test inputs without having to consider computing the corresponding test outputs, reducing the likelihood of making errors.

\subsection{Calculator Agent}
\label{sec:calculator-agent}
Calculator agent is also powered by LLMs.
Based on each test input generated by Designer agent, 
Calculator agent is instructed to determine the corresponding test output and write the entire assert statement.
In Calculator agent, the system prompt defines the conversation format in addition to role assignment, 
and the 1-shot prompt contains a full example of a conversation trajectory. 
Please refer to Figures~\ref{fig:calculator-agent-system-prompt}--\ref{fig:calculator-agent-few-shot-prompt} in
Appendix~\ref{sec:prompt-examples} for the details of the prompts used by Calculator agent.

%When given the function definition, docstring and a test input, Calculator agent will 
%trigger the test output generation.

\textbf{Interaction with a Python Interpreter.}
We create a ReAct format conversation chain for LLMs to interact with the Python interpreter.
The conversation starts with a user message that contains the function definition, docstring and a test input.
Note that if the 1-shot setting is used, the 1-shot prompt should be prefixed to this user message. 
In each round, the LLM will think about what to do and write a Python code snippet or write the test case directly.
The executed result of the Python code snippet will be returned to the LLM subsequently.
The conversation ends when the LLM writes the test case successfully.

During the conversation process, if the LLM only generates thinking text without code and test case, a \emph{go\_on\_prompt} like ``Observation: go on.'' will be sent to the LLM. 
The max round of iterations is set to 5, if the LLM fails to write the test case after 5 rounds, 
a \emph{final\_prompt} like ``Thought: I now know the final answer.\verb|\|nTest Case:'' will be sent to the LLM to force it to write the test case.

Once started, Calculator agent will run a Python shell environment.
During the conversation, all the code snippets will be executed in the same context, which means the subsequent code snippets can access variables from preceding code snippets.

Figure~\ref{fig:conversation} shows an example of the conversation process produced by Calculator agent.
In this example, Calculator agent is tasked with parsing the input string into a list of nested parentheses while disregarding any spaces within the input string.
To handle this question, the LLM divides the task into two steps: (1) Remove any spaces from the input string;
(2) Iterate through the string and separate it into parentheses.
For each step, the LLM writes Python code snippet and obtains the executed results from the Python interpreter.
Finally, the LLM receives the separated parentheses and writes the complete test case.
Please refer to Figures~\ref{fig:example-1-1}--\ref{fig:example-4-8}
in Appendix~\ref{sec:prompt-examples} for more examples.

\begin{table*}[htbp]
\renewcommand{\arraystretch}{1.0}
\centering
\footnotesize
\tabcolsep =6.5pt
\begin{tabular}{llcccccc}
\toprule
\multirow{2}{*}{\textbf{Model}} & \multirow{2}{*}{\textbf{Method}} & \multicolumn{3}{c}{\textbf{HumanEval-no-exp}} && \multicolumn{2}{c}{\textbf{LeetCode-no-exp}} \\
\cmidrule{3-5}
\cmidrule{7-8}
& & \textbf{Accuracy} (\%) & \textbf{Line Cov} (\%) & \textbf{CwB} (\%) && \textbf{Accuracy} (\%) & \textbf{Line Cov} (\%) \\
\midrule
\multirow{2}{*}{GPT-3.5}
& Tester Agent (1-shot)   & 74.02 & 74.69 & 74.15 && 38.97 & 73.66 \\
& TestChain & \textbf{80.85} & \textbf{77.53} & \textbf{80.80} && \textbf{48.72} & \textbf{80.23} \\
\midrule
\multirow{2}{*}{GPT-4}
& Tester Agent (1-shot)   & 84.63 & 77.04 & 83.11 && 57.95 & 88.47 \\
& TestChain & \textbf{90.24} & \textbf{80.00} & \textbf{88.66} && \textbf{71.79} & \textbf{90.60} \\
\bottomrule
\end{tabular}
\caption{
Evaluation results of the Tester Agent with 1-shot prompt approach and the TestChain approach.
}
\label{tab:table2}
\end{table*}

\section{Evaluation of TestChain}

\subsection{Experimental Setup}

We follow the same setups in Section~\ref{sec:can-experimental-setup} for datasets and metrics.
We consider two advanced models GPT-3.5 and GPT-4 in our TestChain framework.
Because TestChain uses the 1-shot setting by default, we choose 
the Test Agent (1-shot) method depicted in Section~\ref{sec:can-experimental-setup}
as the baseline for comparison.

%We adapt two stronger models GPT-3.5 and GPT-4 to the TestChain framework and 
%consider the few-shot method as the baseline.
%For the designer agent and calculator agent, we create system prompt, few-shot prompt, and generation instruction similar to the prompts in Section~\ref{sec:can-experimental-setup}.

\subsection{Evaluation Results}
\label{sec:evaluation-results}

%We conduct statistics on the number of successfully generated test cases in Figure~\ref{fig:generated-2}.
%Experiments demonstrate that for stronger models like GPT-3.5 and GPT-4, both the few-shot method and the TestChain framework can generate enough efficient test cases.
%Thus the key to influencing the accuracy of test cases lies in whether the prediction output results are correct.

\begin{figure}[htbp]
\begin{center}
%\vskip 0.2in
\centering
\includegraphics[width=\linewidth]{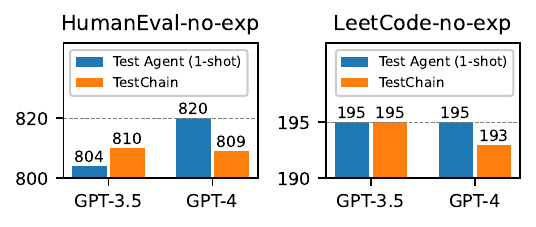}
\caption{
Statistics on the number of test cases successfully generated.
}
\label{fig:generated-2}
\end{center}
%\vskip -0.2in
\end{figure}

We analyze the statistics regarding the number of successfully generated test cases, as illustrated in Figure~\ref{fig:generated-2}. 
Our results indicate that the advanced models such as GPT-3.5 and GPT-4 exhibit the strong capability to produce a sufficient number of valid test cases using both the Test Agent (1-shot) method and the TestChain method. 
Hence, the accuracy comparison among generation methods primarily relies on the number of correct test cases that a method can generate.

\begin{figure*}[htbp]
\begin{center}
\begin{subfigure}{1\columnwidth}
\vskip 0.2in
\centering
\includegraphics[width=\linewidth]{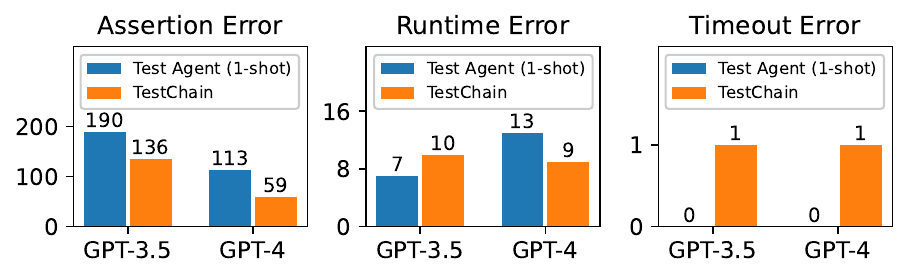}
\caption{
On the HumanEval-no-exp dataset.
}
\vskip -0.2in
\end{subfigure}
\begin{subfigure}{\columnwidth}
\vskip 0.2in
\centering
\includegraphics[width=\linewidth]{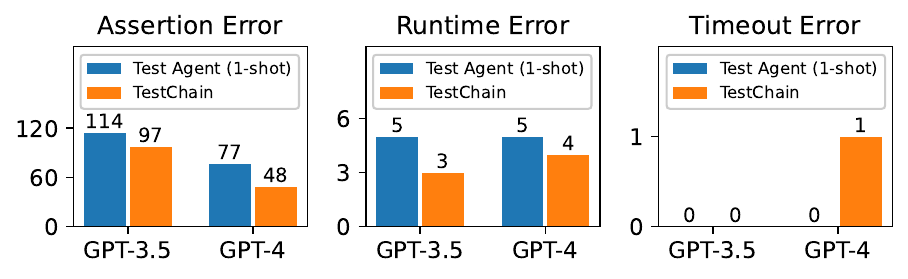}
\caption{
On the LeetCode-no-exp dataset.
}
\vskip -0.2in
\end{subfigure}
\caption{
Statistics on the number of incorrect test cases for each type of error.
}
\label{fig:counts-2}
\end{center}
\end{figure*}

%Table~\ref{tab:table2} shows the comparison of all the metrics for the few-shot method and the TestChain framework.
%We find that compared to the baseline, the TestChain framework has achieved significant improvements across all metrics and datasets.
%We observe that on the more challenging dataset LeetCode-no-exp with GPT-4, the accuracy of the TestChain framework outperforms the baseline by \textbf{13.84\%}.
%For other metrics such as Line Cov and CwB, the TestChain framework also achieves state-of-the-art results.
%This demonstrates that the TestChain framework is capable of generating test cases that are both correct and of high quality.

Table~\ref{tab:table2} presents a comparison of all metrics between the TestChain method and the Test Agent (1-shot) method.
It is evident that TestChain achieves significant improvements over the Test Agent (1-shot) baseline across all metrics and datasets, 
regardless of whether GPT-3.5 or GPT-4 is utilized as the backbone model.
Notably, on the challenging LeetCode-no-exp dataset, TestChain with GPT-4 
surpasses the corresponding baseline by 13.84\% in terms of accuracy. 
As for Line Cov and CwB metrics, 
our TestChain method also always achieves better results than the Test Agent (1-shot) method. 
These findings underscore the ability of our proposed TestChain to generate high-quality test cases that are both accurate and reliable.

Figure~\ref{fig:counts-2} displays the statistical distribution of incorrect test cases across different error categories.
It can be seen that Assertion Error is still the most common type of error for all the methods. 
But the number of assertion errors by TestChain is greatly reduced compared to the Test Agent (1-shot) method,
contributing most to TestChain's higher accuracy. 
This demonstrates that our TestChain, 
leveraging Python codes for step-by-step computation and reasoning, 
effectively decreases the incidence of errors in test outputs, consequently improving the accuracy of the test cases generated.

%Figure~\ref{fig:counts-2} illustrates the statistical count of test cases for various types of errors.
%We discovered that the error type with the greatest impact on the accuracy of the generated results is also the Assertion Error.
%For the most prevalent type of error, the Assertion Error, our method achieves a notable reduction across all models and datasets.
%This proves that our method, which relies on Python codes for complex calculations, has reduced the proportion of computational errors in test case outputs, thereby enhancing the accuracy of generated test cases.

%\subsection{How does the Python Interpreter Perform?}

\subsection{Effectiveness of the Ingredients in TestChain}

In this subsection, we want to examine the effectiveness of two ingredients in the TestChain framework:
the decoupling of test input and output generation, and the interaction with the Python interpreter.
To this end, we design a modified TestChain method, where
Calculator agent no longer uses the Python interpreter and instead uses a 1-shot prompt to generate the test case directly.
Please refer to Figures~\ref{fig:calculator-agent-no-py-system-prompt}--\ref{fig:calculator-agent-no-py-few-shot-prompt} in
Appendix~\ref{sec:prompt-examples} for the details of the prompts used by this modified TestChain method.
We use the same setups with Section~\ref{sec:evaluation-results} and compare it with the original TestChain method along with
the Test Agent (1-shot) baseline method. 

\begin{table}[!htbp]
\renewcommand{\arraystretch}{1.0}
\centering
\footnotesize
\tabcolsep =6pt
\begin{tabular}{lcc}
\toprule
\textbf{Method}    & \textbf{Accuracy} (\%) & \textbf{Line Cov} (\%)\\
\midrule
Test Agent (1-shot) & 57.95 & 88.47 \\
TestChain (no-py)  & 61.54 & 83.65 \\
TestChain          & \textbf{71.79} & \textbf{90.60} \\
\bottomrule
\end{tabular}
\caption{
Results on the LeetCode-no-exp dataset with GPT-4.
TestChain (no-py) refers to the TestChain method without the Python interpreter interaction.
}
\label{tab:table3}
\end{table}

Table~\ref{tab:table3} shows the comparison results. 
The main difference between the modified TestChain method and the Test Agent (1-shot) method is that the modified TestChain 
separates the generation of test inputs and test outputs, while the Test Agent method generates them together.
The modified TestChain outperforms the Test Agent (1-shot) method by 3.59\% in accuracy, demonstrating the benefit of the decoupling through two agents. 
The original TestChain achieves substantial improvements over the modified one in terms of 
both accuracy and line coverage. This suggests that in TestChain, the interaction with the Python interpreter 
is vital in enhancing the generation performance.

\section{Conclusion}

In this paper, we have conducted a thorough performance evaluation of LLMs in test case generation. 
We find that, for relatively hard questions such as those in LeetCode-hard,
even the state-of-the-art LLMs like GPT-4 have great difficulty in generating correct test cases.
To address this challenge, we propose TestChain, characterized by decoupling the generation of test inputs and test outputs,
and interacting with the Python interpreter through the ReAct chain. 
Our results clearly demonstrate the superiority of TestChain over the baselines.

\section*{Limitations}

The proposed TestChain framework demands high model capabilities, thus it is tailored for more robust models such as GPT-3.5 and GPT-4. 
Future work can explore how to enhance the performance of weaker models in test case generation based on the TestChain paradigm.

% \section*{Acknowledgements}

\bibliography{anthology,custom}

\begin{thebibliography}{28}
\expandafter\ifx\csname natexlab\endcsname\relax\def\natexlab#1{#1}\fi

\bibitem[{Achiam et~al.(2023)Achiam, Adler, Agarwal, Ahmad, Akkaya, Aleman, Almeida, Altenschmidt, Altman, Anadkat et~al.}]{achiam2023gpt}
Josh Achiam, Steven Adler, Sandhini Agarwal, Lama Ahmad, Ilge Akkaya, Florencia~Leoni Aleman, Diogo Almeida, Janko Altenschmidt, Sam Altman, Shyamal Anadkat, et~al. 2023.
\newblock Gpt-4 technical report.
\newblock \emph{arXiv preprint arXiv:2303.08774}.

\bibitem[{Andrews et~al.(2006)Andrews, Briand, Labiche, and Namin}]{andrews2006using}
James~H Andrews, Lionel~C Briand, Yvan Labiche, and Akbar~Siami Namin. 2006.
\newblock Using mutation analysis for assessing and comparing testing coverage criteria.
\newblock \emph{IEEE Transactions on Software Engineering}, 32(8):608--624.

\bibitem[{Anil et~al.(2023)Anil, Dai, Firat, Johnson, Lepikhin, Passos, Shakeri, Taropa, Bailey, Chen et~al.}]{anil2023palm}
Rohan Anil, Andrew~M Dai, Orhan Firat, Melvin Johnson, Dmitry Lepikhin, Alexandre Passos, Siamak Shakeri, Emanuel Taropa, Paige Bailey, Zhifeng Chen, et~al. 2023.
\newblock Palm 2 technical report.
\newblock \emph{arXiv preprint arXiv:2305.10403}.

\bibitem[{Chen et~al.(2022)Chen, Zhang, Nguyen, Zan, Lin, Lou, and Chen}]{chen2022codet}
Bei Chen, Fengji Zhang, Anh Nguyen, Daoguang Zan, Zeqi Lin, Jian-Guang Lou, and Weizhu Chen. 2022.
\newblock Codet: Code generation with generated tests.
\newblock \emph{arXiv preprint arXiv:2207.10397}.

\bibitem[{Chen et~al.(2021)Chen, Tworek, Jun, Yuan, Pinto, Kaplan, Edwards, Burda, Joseph, Brockman et~al.}]{chen2021evaluating}
Mark Chen, Jerry Tworek, Heewoo Jun, Qiming Yuan, Henrique Ponde de~Oliveira Pinto, Jared Kaplan, Harri Edwards, Yuri Burda, Nicholas Joseph, Greg Brockman, et~al. 2021.
\newblock Evaluating large language models trained on code.
\newblock \emph{arXiv preprint arXiv:2107.03374}.

\bibitem[{Dong et~al.(2023)Dong, Jiang, Jin, and Li}]{dong2023self}
Yihong Dong, Xue Jiang, Zhi Jin, and Ge~Li. 2023.
\newblock Self-collaboration code generation via chatgpt.
\newblock \emph{arXiv preprint arXiv:2304.07590}.

\bibitem[{Frieder et~al.(2023)Frieder, Pinchetti, Chevalier, Griffiths, Salvatori, Lukasiewicz, Petersen, and Berner}]{frieder2023mathematical}
Simon Frieder, Luca Pinchetti, Alexis Chevalier, Ryan-Rhys Griffiths, Tommaso Salvatori, Thomas Lukasiewicz, Philipp~Christian Petersen, and Julius Berner. 2023.
\newblock Mathematical capabilities of chatgpt.
\newblock \emph{arXiv preprint arXiv:2301.13867}.

\bibitem[{Huang et~al.(2023{\natexlab{a}})Huang, Bu, and Cui}]{huang2023codecot}
Dong Huang, Qingwen Bu, and Heming Cui. 2023{\natexlab{a}}.
\newblock Codecot and beyond: Learning to program and test like a developer.
\newblock \emph{arXiv preprint arXiv:2308.08784}.

\bibitem[{Huang et~al.(2023{\natexlab{b}})Huang, Bu, Zhang, Luck, and Cui}]{huang2023agentcoder}
Dong Huang, Qingwen Bu, Jie~M Zhang, Michael Luck, and Heming Cui. 2023{\natexlab{b}}.
\newblock Agentcoder: Multi-agent-based code generation with iterative testing and optimisation.
\newblock \emph{arXiv preprint arXiv:2312.13010}.

\bibitem[{Kang et~al.(2023)Kang, Yoon, and Yoo}]{kang2023large}
Sungmin Kang, Juyeon Yoon, and Shin Yoo. 2023.
\newblock Large language models are few-shot testers: Exploring llm-based general bug reproduction.
\newblock In \emph{2023 IEEE/ACM 45th International Conference on Software Engineering (ICSE)}, pages 2312--2323. IEEE.

\bibitem[{Li et~al.(2023)Li, Allal, Zi, Muennighoff, Kocetkov, Mou, Marone, Akiki, Li, Chim et~al.}]{li2023starcoder}
Raymond Li, Loubna~Ben Allal, Yangtian Zi, Niklas Muennighoff, Denis Kocetkov, Chenghao Mou, Marc Marone, Christopher Akiki, Jia Li, Jenny Chim, et~al. 2023.
\newblock Starcoder: may the source be with you!
\newblock \emph{arXiv preprint arXiv:2305.06161}.

\bibitem[{Li et~al.(2022)Li, Choi, Chung, Kushman, Schrittwieser, Leblond, Eccles, Keeling, Gimeno, Dal~Lago et~al.}]{li2022competition}
Yujia Li, David Choi, Junyoung Chung, Nate Kushman, Julian Schrittwieser, R{\'e}mi Leblond, Tom Eccles, James Keeling, Felix Gimeno, Agustin Dal~Lago, et~al. 2022.
\newblock Competition-level code generation with alphacode.
\newblock \emph{Science}, 378(6624):1092--1097.

\bibitem[{Liu et~al.(2023)Liu, Ning, Teng, Liu, Zhou, and Zhang}]{liu2023evaluating}
Hanmeng Liu, Ruoxi Ning, Zhiyang Teng, Jian Liu, Qiji Zhou, and Yue Zhang. 2023.
\newblock Evaluating the logical reasoning ability of chatgpt and gpt-4.
\newblock \emph{arXiv preprint arXiv:2304.03439}.

\bibitem[{Nakano et~al.(2021)Nakano, Hilton, Balaji, Wu, Ouyang, Kim, Hesse, Jain, Kosaraju, Saunders et~al.}]{nakano2021webgpt}
Reiichiro Nakano, Jacob Hilton, Suchir Balaji, Jeff Wu, Long Ouyang, Christina Kim, Christopher Hesse, Shantanu Jain, Vineet Kosaraju, William Saunders, et~al. 2021.
\newblock Webgpt: Browser-assisted question-answering with human feedback.
\newblock \emph{arXiv preprint arXiv:2112.09332}.

\bibitem[{Nijkamp et~al.(2022)Nijkamp, Pang, Hayashi, Tu, Wang, Zhou, Savarese, and Xiong}]{nijkamp2022codegen}
Erik Nijkamp, Bo~Pang, Hiroaki Hayashi, Lifu Tu, Huan Wang, Yingbo Zhou, Silvio Savarese, and Caiming Xiong. 2022.
\newblock Codegen: An open large language model for code with multi-turn program synthesis.
\newblock \emph{arXiv preprint arXiv:2203.13474}.

\bibitem[{Romera-Paredes et~al.(2023)Romera-Paredes, Barekatain, Novikov, Balog, Kumar, Dupont, Ruiz, Ellenberg, Wang, Fawzi, Kohli, and Fawzi}]{FunSearch2023}
Bernardino Romera-Paredes, Mohammadamin Barekatain, Alexander Novikov, Matej Balog, M.~Pawan Kumar, Emilien Dupont, Francisco J.~R. Ruiz, Jordan Ellenberg, Pengming Wang, Omar Fawzi, Pushmeet Kohli, and Alhussein Fawzi. 2023.
\newblock \href {https://doi.org/10.1038/s41586-023-06924-6} {Mathematical discoveries from program search with large language models}.
\newblock \emph{Nature}.

\bibitem[{Roziere et~al.(2023)Roziere, Gehring, Gloeckle, Sootla, Gat, Tan, Adi, Liu, Remez, Rapin et~al.}]{roziere2023code}
Baptiste Roziere, Jonas Gehring, Fabian Gloeckle, Sten Sootla, Itai Gat, Xiaoqing~Ellen Tan, Yossi Adi, Jingyu Liu, Tal Remez, J{\'e}r{\'e}my Rapin, et~al. 2023.
\newblock Code llama: Open foundation models for code.
\newblock \emph{arXiv preprint arXiv:2308.12950}.

\bibitem[{Sch{\"a}fer et~al.(2023)Sch{\"a}fer, Nadi, Eghbali, and Tip}]{schafer2023empirical}
Max Sch{\"a}fer, Sarah Nadi, Aryaz Eghbali, and Frank Tip. 2023.
\newblock An empirical evaluation of using large language models for automated unit test generation.
\newblock \emph{IEEE Transactions on Software Engineering}.

\bibitem[{Shen et~al.(2023)Shen, Song, Tan, Li, Lu, and Zhuang}]{shen2023hugginggpt}
Yongliang Shen, Kaitao Song, Xu~Tan, Dongsheng Li, Weiming Lu, and Yueting Zhuang. 2023.
\newblock Hugginggpt: Solving ai tasks with chatgpt and its friends in huggingface.
\newblock \emph{arXiv preprint arXiv:2303.17580}.

\bibitem[{Shinn et~al.(2023)Shinn, Cassano, Gopinath, Narasimhan, and Yao}]{shinn2023reflexion}
Noah Shinn, Federico Cassano, Ashwin Gopinath, Karthik~R Narasimhan, and Shunyu Yao. 2023.
\newblock Reflexion: Language agents with verbal reinforcement learning.
\newblock In \emph{Thirty-seventh Conference on Neural Information Processing Systems}.

\bibitem[{Tufano et~al.(2020)Tufano, Drain, Svyatkovskiy, Deng, and Sundaresan}]{tufano2020unit}
Michele Tufano, Dawn Drain, Alexey Svyatkovskiy, Shao~Kun Deng, and Neel Sundaresan. 2020.
\newblock Unit test case generation with transformers and focal context.
\newblock \emph{arXiv preprint arXiv:2009.05617}.

\bibitem[{Wang et~al.(2023{\natexlab{a}})Wang, Liu, and Awadallah}]{wang2023EcoOptiGen}
Chi Wang, Susan~Xueqing Liu, and Ahmed~H. Awadallah. 2023{\natexlab{a}}.
\newblock Cost-effective hyperparameter optimization for large language model generation inference.
\newblock In \emph{AutoML'23}.

\bibitem[{Wang et~al.(2023{\natexlab{b}})Wang, Liu, Wang, Cui, Ding, Liu, and Yu}]{wang2023intervenor}
Hanbin Wang, Zhenghao Liu, Shuo Wang, Ganqu Cui, Ning Ding, Zhiyuan Liu, and Ge~Yu. 2023{\natexlab{b}}.
\newblock Intervenor: Prompt the coding ability of large language models with the interactive chain of repairing.
\newblock \emph{arXiv preprint arXiv:2311.09868}.

\bibitem[{Wolf et~al.(2019)Wolf, Debut, Sanh, Chaumond, Delangue, Moi, Cistac, Rault, Louf, Funtowicz et~al.}]{wolf2019huggingface}
Thomas Wolf, Lysandre Debut, Victor Sanh, Julien Chaumond, Clement Delangue, Anthony Moi, Pierric Cistac, Tim Rault, R{\'e}mi Louf, Morgan Funtowicz, et~al. 2019.
\newblock Huggingface's transformers: State-of-the-art natural language processing.
\newblock \emph{arXiv preprint arXiv:1910.03771}.

\bibitem[{Wu et~al.(2023{\natexlab{a}})Wu, Bansal, Zhang, Wu, Zhang, Zhu, Li, Jiang, Zhang, and Wang}]{wu2023autogen}
Qingyun Wu, Gagan Bansal, Jieyu Zhang, Yiran Wu, Shaokun Zhang, Erkang Zhu, Beibin Li, Li~Jiang, Xiaoyun Zhang, and Chi Wang. 2023{\natexlab{a}}.
\newblock Autogen: Enabling next-gen llm applications via multi-agent conversation framework.
\newblock \emph{arXiv preprint arXiv:2308.08155}.

\bibitem[{Wu et~al.(2023{\natexlab{b}})Wu, Jia, Zhang, Li, Zhu, Wang, Lee, Peng, Wu, and Wang}]{wu2023empirical}
Yiran Wu, Feiran Jia, Shaokun Zhang, Hangyu Li, Erkang Zhu, Yue Wang, Yin~Tat Lee, Richard Peng, Qingyun Wu, and Chi Wang. 2023{\natexlab{b}}.
\newblock An empirical study on challenging math problem solving with gpt-4.
\newblock In \emph{ArXiv preprint arXiv:2306.01337}.

\bibitem[{Yao et~al.(2022)Yao, Zhao, Yu, Du, Shafran, Narasimhan, and Cao}]{yao2022react}
Shunyu Yao, Jeffrey Zhao, Dian Yu, Nan Du, Izhak Shafran, Karthik Narasimhan, and Yuan Cao. 2022.
\newblock React: Synergizing reasoning and acting in language models.
\newblock \emph{arXiv preprint arXiv:2210.03629}.

\bibitem[{Zhang et~al.(2023)Zhang, Song, Ji, Meng et~al.}]{zhang2023well}
Ying Zhang, Wenjia Song, Zhengjie Ji, Na~Meng, et~al. 2023.
\newblock How well does llm generate security tests?
\newblock \emph{arXiv preprint arXiv:2310.00710}.

\end{thebibliography}
\bibliographystyle{acl_natbib}

\appendix

\section{Dataset Examples}
\label{sec:dataset-examples}

\begin{figure*}[!htbp]
\vskip 0.2in
\begin{center}
\centerline{\includegraphics[width=2\columnwidth]{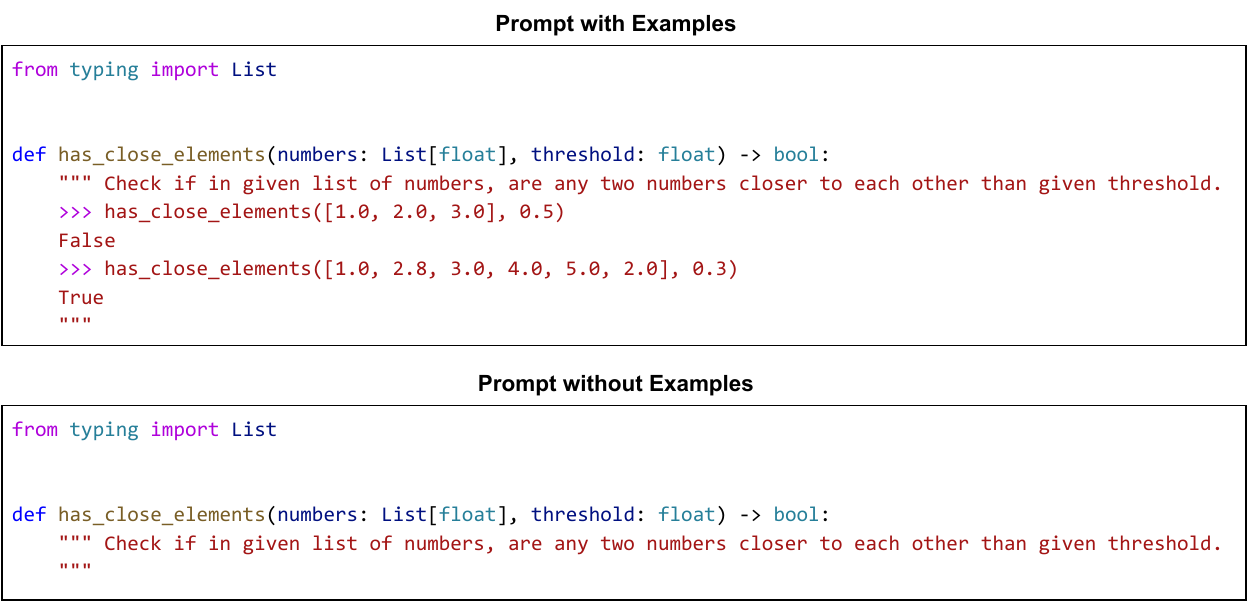}}
\caption{
Illustration of the modifications for the prompt of the HumanEval and LeetCode-hard datasets.
}
\label{fig:no-examples}
\end{center}
\vskip -0.2in
\end{figure*}

Figure~\ref{fig:no-examples} shows the prompt of HumanEval and LeetCode datasets with and without examples.

\section{Prompts for Test Case Generation}
\label{sec:prompt-examples}

\begin{figure*}[!htbp]
\vskip 0.2in
\begin{center}
\centerline{\includegraphics[width=2\columnwidth]{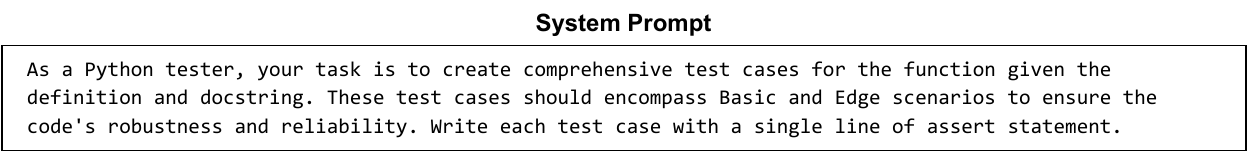}}
\caption{
System prompt for the Test Agent method.
}
\label{fig:system-prompt}
\end{center}
\vskip -0.2in
\end{figure*}

\begin{figure*}[!htbp]
\vskip 0.2in
\begin{center}
\centerline{\includegraphics[width=2\columnwidth]{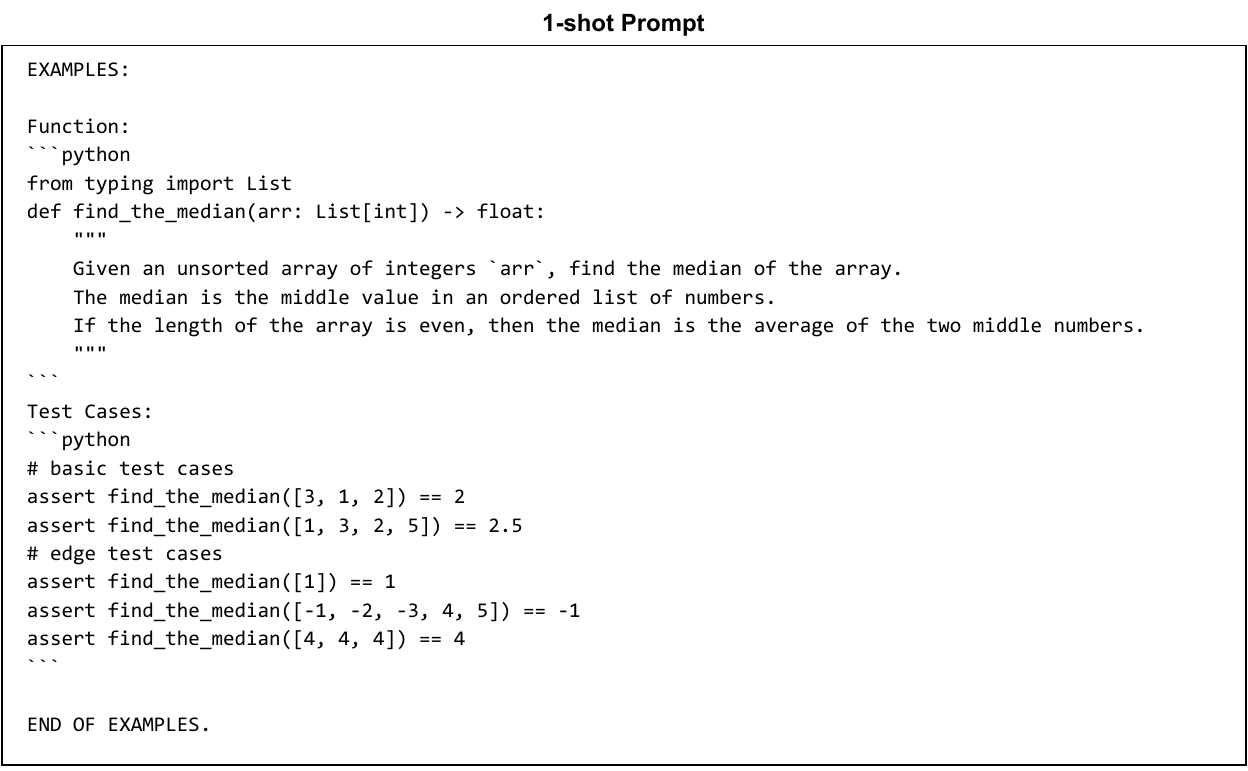}}
\caption{
1-shot Prompt for the Test Agent method.
}
\label{fig:few-shot-prompt}
\end{center}
\vskip -0.2in
\end{figure*}

\begin{figure*}[!htbp]
\vskip 0.2in
\begin{center}
\centerline{\includegraphics[width=2\columnwidth]{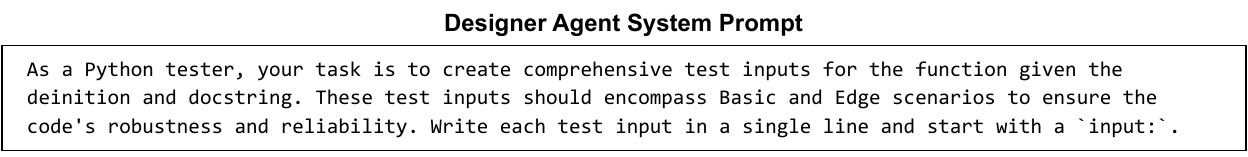}}
\caption{
System prompt for the Designer agent.
}
\label{fig:designer-agent-system-prompt}
\end{center}
\vskip -0.2in
\end{figure*}

\begin{figure*}[!htbp]
\vskip 0.2in
\begin{center}
\centerline{\includegraphics[width=2\columnwidth]{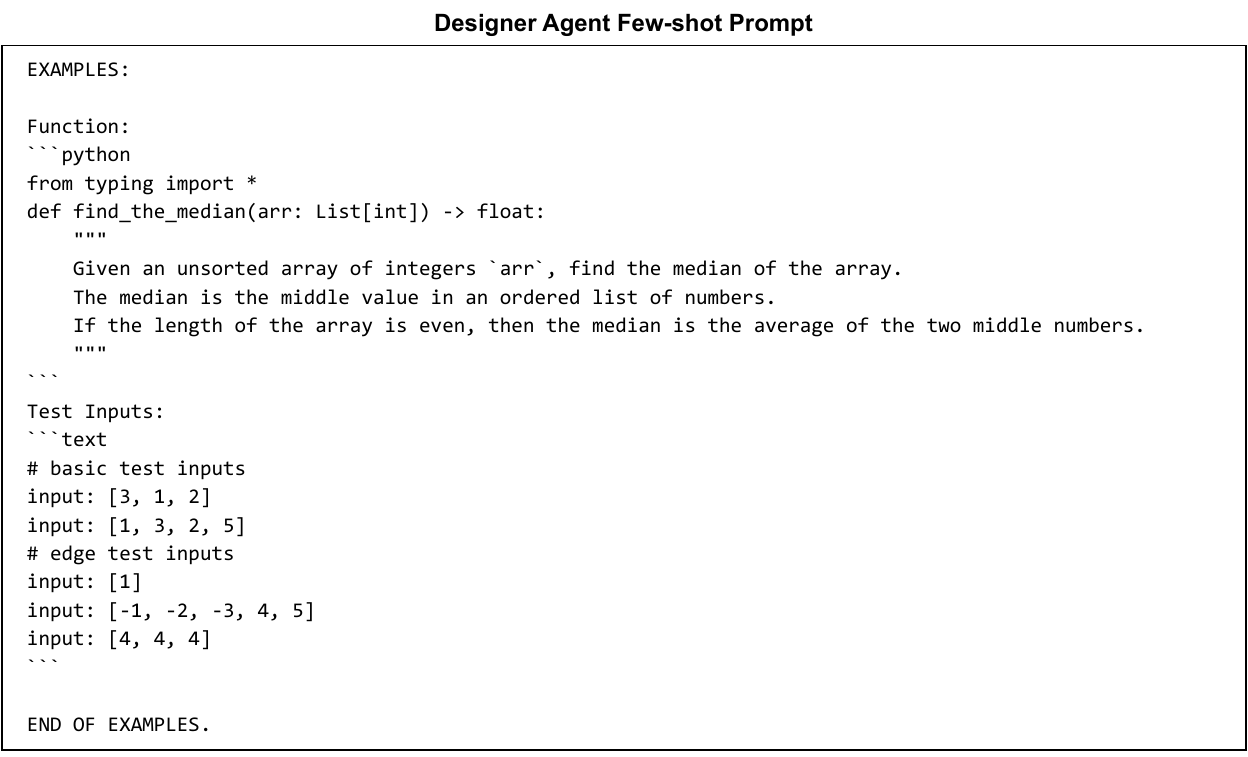}}
\caption{
1-shot prompt for the Designer agent.
}
\label{fig:designer-agent-few-shot-prompt}
\end{center}
\vskip -0.2in
\end{figure*}

\begin{figure*}[!htbp]
\vskip 0.2in
\begin{center}
\centerline{\includegraphics[width=2\columnwidth]{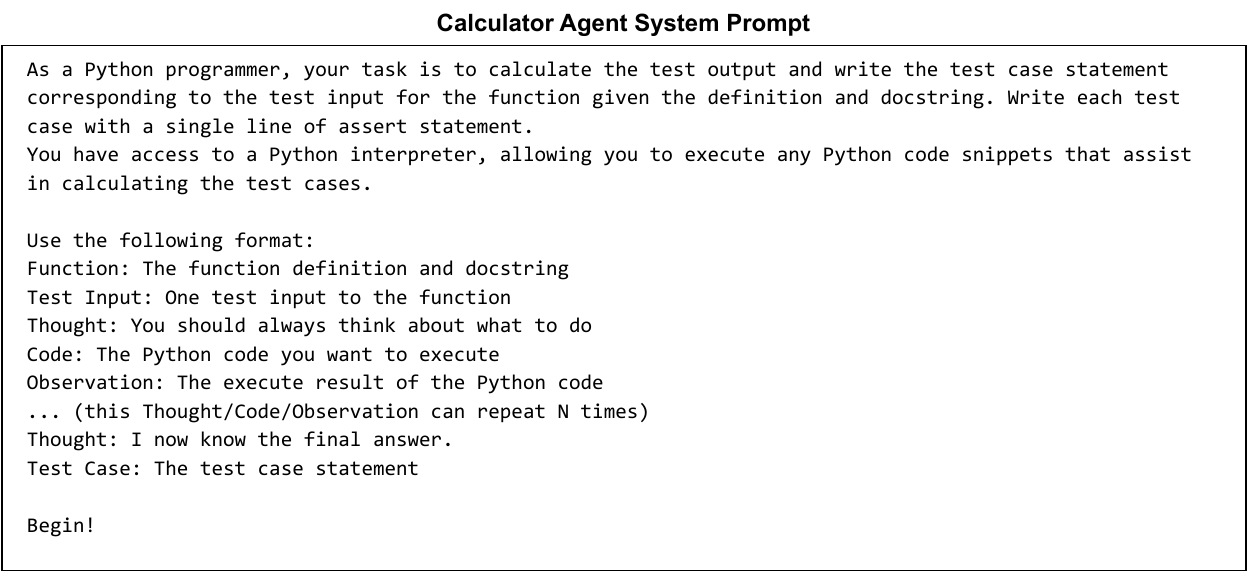}}
\caption{
System prompt for the Calculator agent.
}
\label{fig:calculator-agent-system-prompt}
\end{center}
\vskip -0.2in
\end{figure*}

\begin{figure*}[!htbp]
\vskip 0.2in
\begin{center}
\centerline{\includegraphics[width=2\columnwidth]{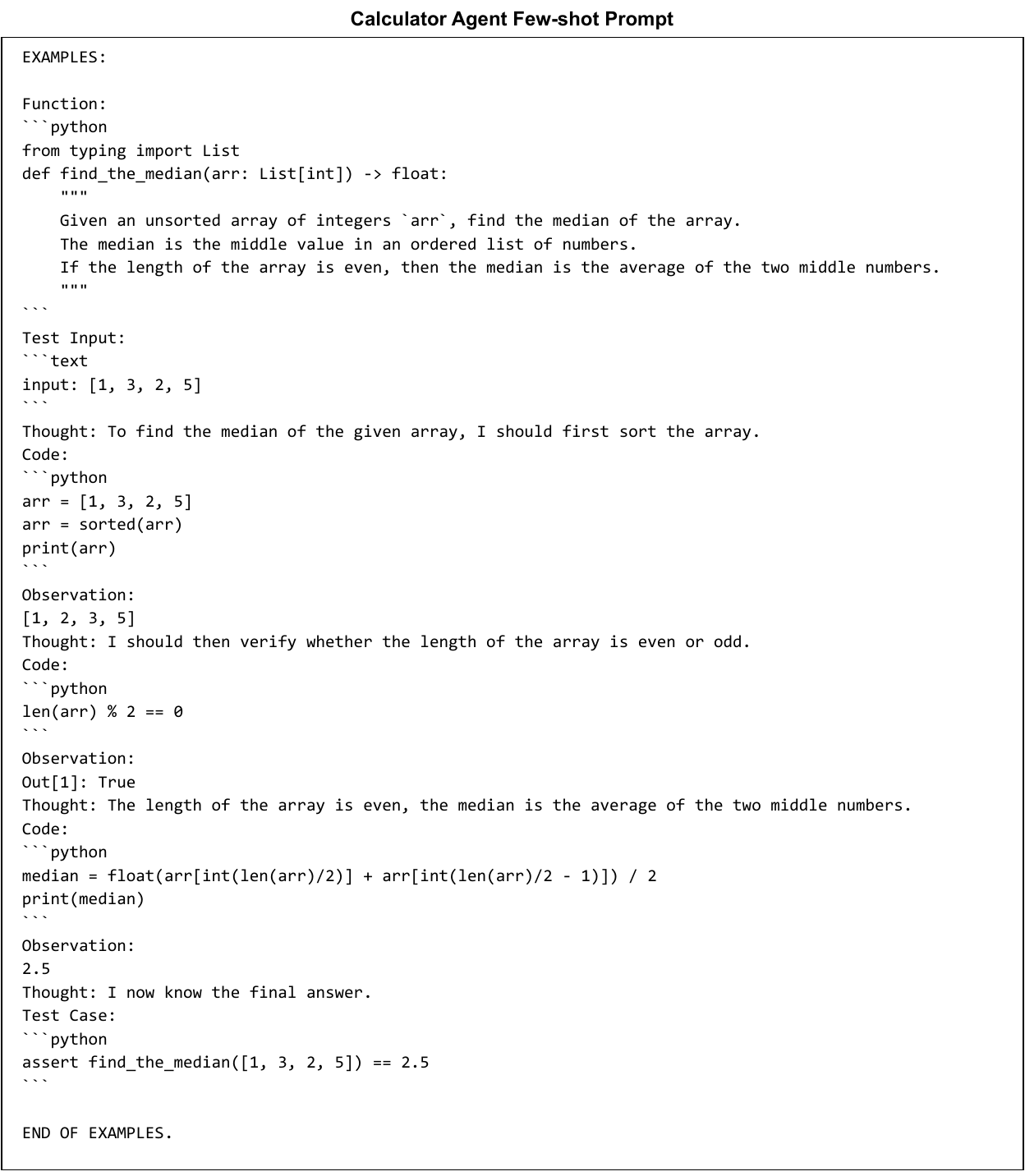}}
\caption{
1-shot prompt for the Calculator agent.
}
\label{fig:calculator-agent-few-shot-prompt}
\end{center}
\vskip -0.2in
\end{figure*}

\begin{figure*}[!htbp]
\vskip 0.2in
\begin{center}
\centerline{\includegraphics[width=2\columnwidth]{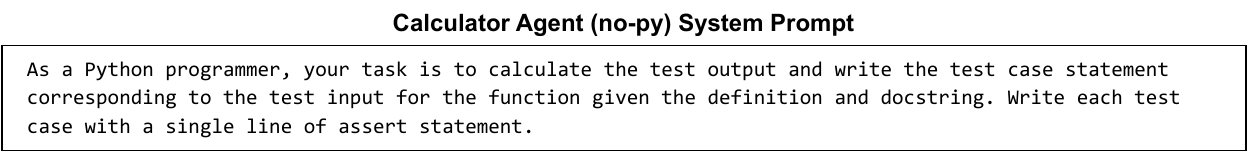}}
\caption{
System prompt for the Calculator agent without the Python interpreter.
}
\label{fig:calculator-agent-no-py-system-prompt}
\end{center}
\vskip -0.2in
\end{figure*}

\begin{figure*}[!htbp]
\vskip 0.2in
\begin{center}
\centerline{\includegraphics[width=2\columnwidth]{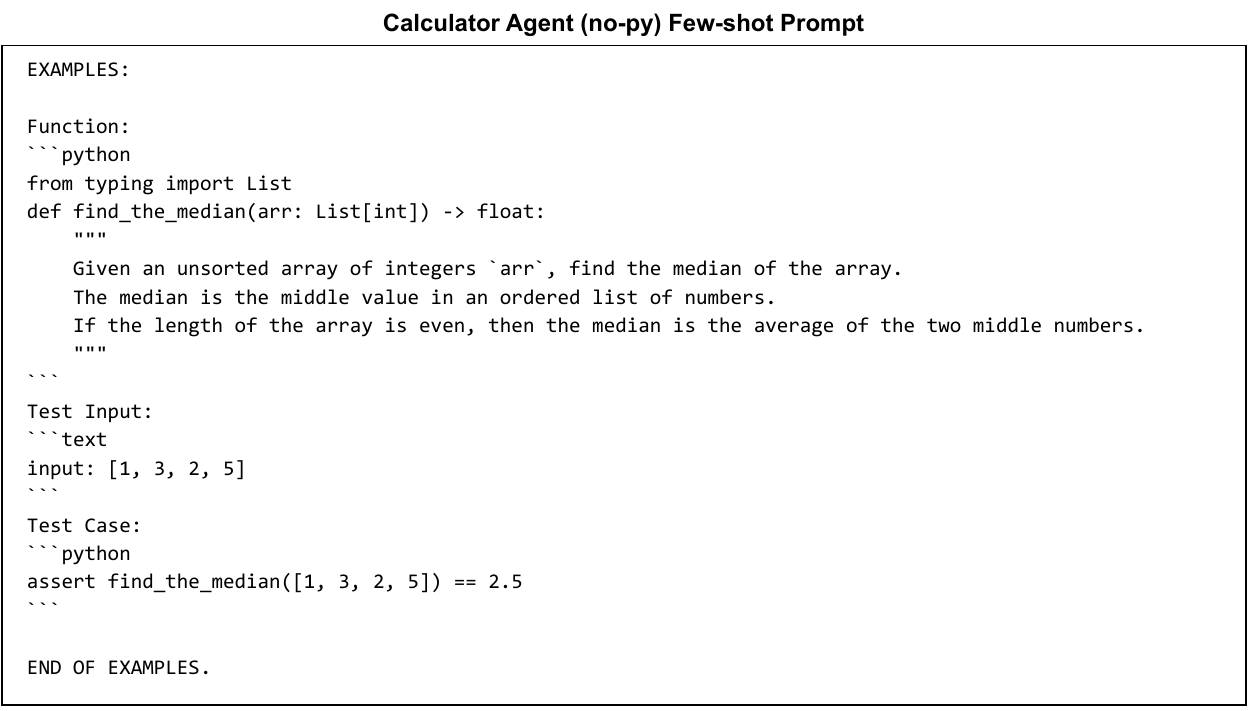}}
\caption{
1-shot prompt for the Calculator agent without the Python interpreter.
}
\label{fig:calculator-agent-no-py-few-shot-prompt}
\end{center}
\vskip -0.2in
\end{figure*}

Figures~\ref{fig:system-prompt}--\ref{fig:calculator-agent-no-py-few-shot-prompt}
provide prompts used by different generation methods considered in our paper. 

%These are prompts we used in the experiments.

\section{Examples of Trajectories by Calculator Agent}
\label{sec:successfully-generated-examples}

\begin{figure*}[ht]
\textbf{Example 1.}
\vskip 0.2in
\begin{center}
\centerline{\includegraphics[width=2\columnwidth]{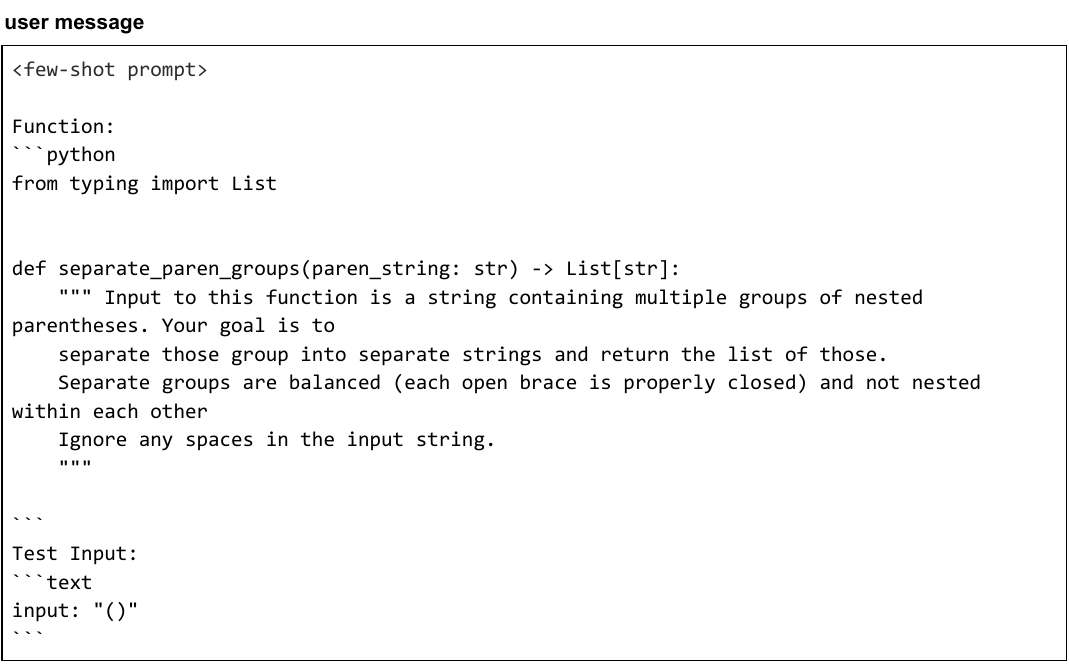}}
\caption{
Message 1 of Example 1.
}
\label{fig:example-1-1}
\end{center}
\vskip -0.2in
\end{figure*}

\begin{figure*}[ht]
\vskip 0.2in
\begin{center}
\centerline{\includegraphics[width=2\columnwidth]{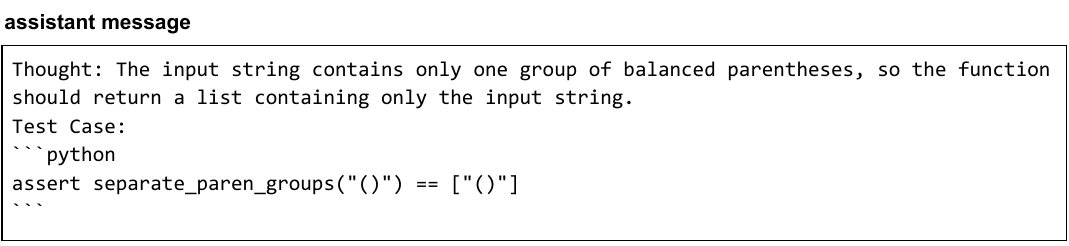}}
\caption{
Message 2 of example 1.
}
\label{fig:example-1-2}
\end{center}
\vskip -0.2in
\end{figure*}

\begin{figure*}[ht]
\textbf{Example 2.}
\vskip 0.2in
\begin{center}
\centerline{\includegraphics[width=2\columnwidth]{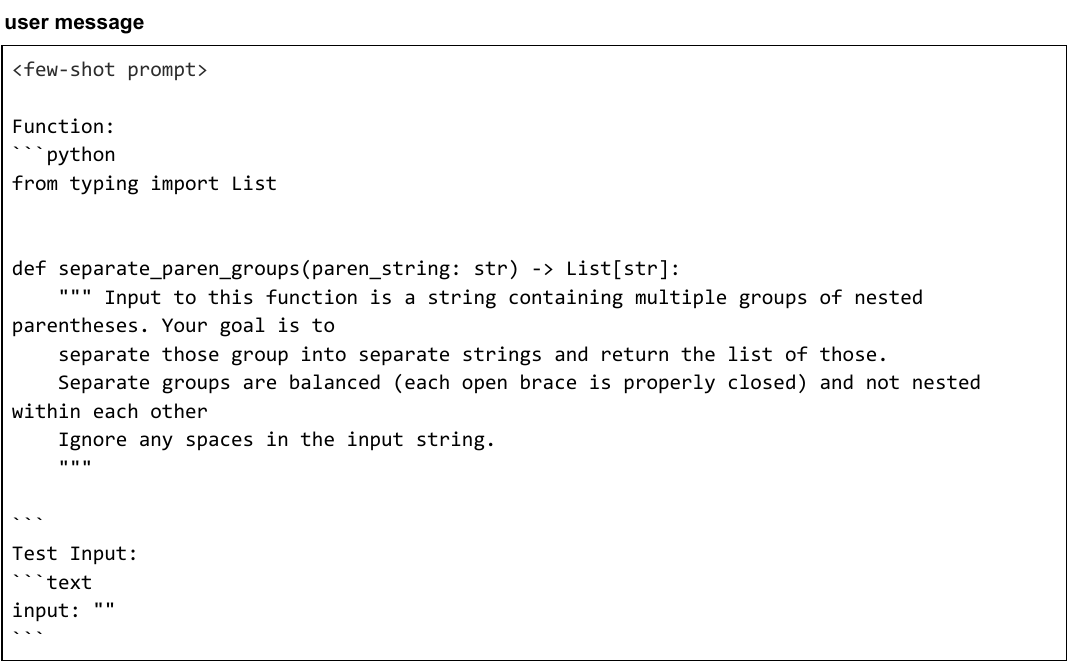}}
\caption{
Message 1 of Example 2.
}
\label{fig:example-2-1}
\end{center}
\vskip -0.2in
\end{figure*}

\begin{figure*}[ht]
\vskip 0.2in
\begin{center}
\centerline{\includegraphics[width=2\columnwidth]{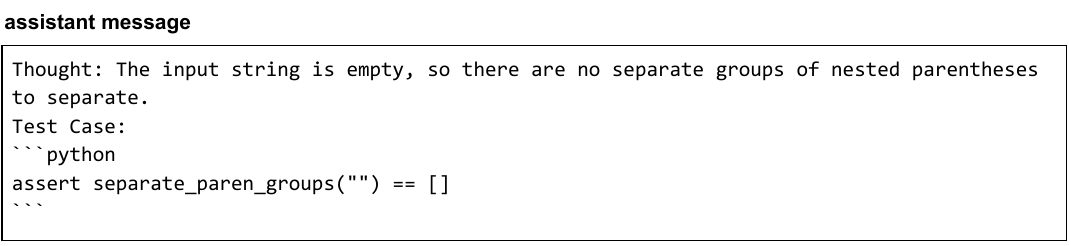}}
\caption{
Message 2 of Example 2.
}
\label{fig:example-2-2}
\end{center}
\vskip -0.2in
\end{figure*}

\begin{figure*}[ht]
\textbf{Example 3.}
\vskip 0.2in
\begin{center}
\centerline{\includegraphics[width=2\columnwidth]{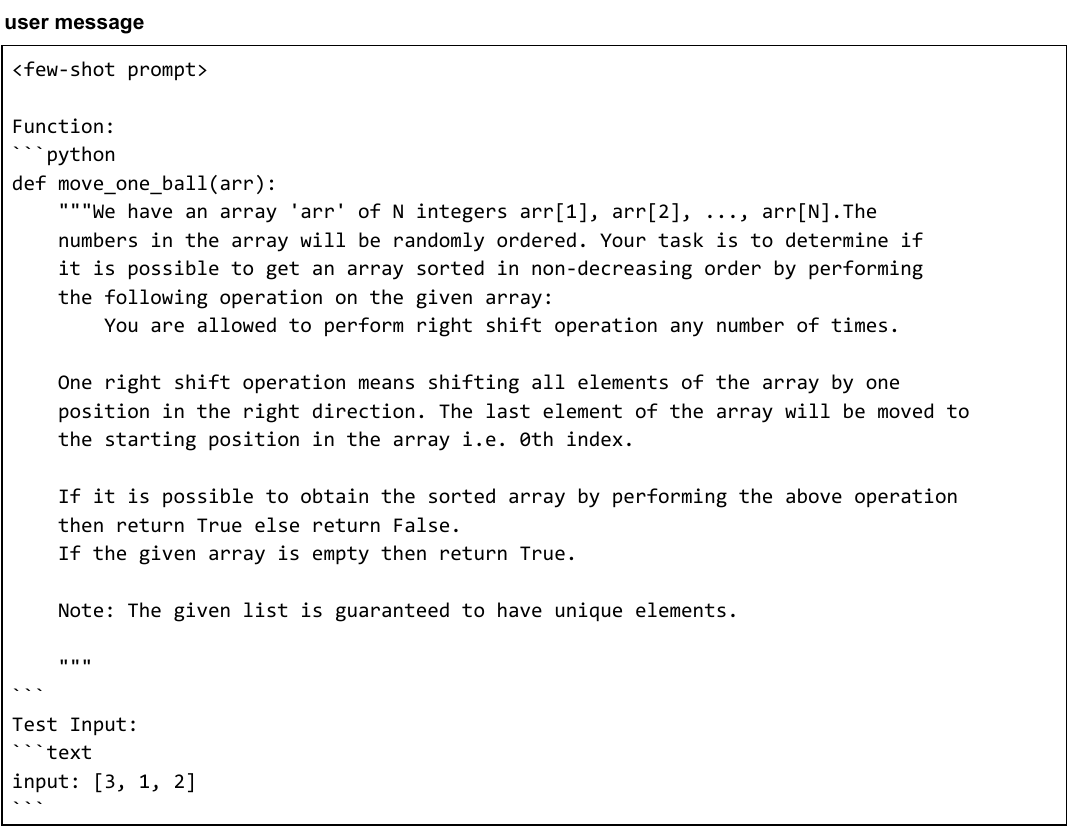}}
\caption{
Message 1 of Example 3.
}
\label{fig:example-3-1}
\end{center}
\vskip -0.2in
\end{figure*}

\begin{figure*}[ht]
\vskip 0.2in
\begin{center}
\centerline{\includegraphics[width=2\columnwidth]{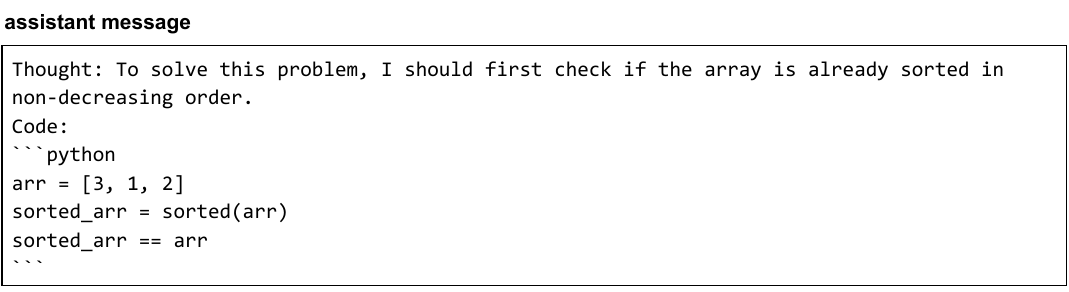}}
\caption{
Message 2 of Example 3.
}
\label{fig:example-3-2}
\end{center}
\vskip -0.2in
\end{figure*}

\begin{figure*}[ht]
\vskip 0.2in
\begin{center}
\centerline{\includegraphics[width=2\columnwidth]{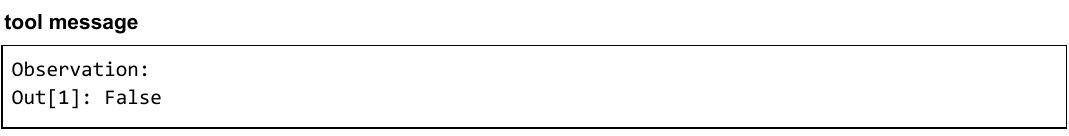}}
\caption{
Message 3 of Example 3.
}
\label{fig:example-3-3}
\end{center}
\vskip -0.2in
\end{figure*}

\begin{figure*}[ht]
\vskip 0.2in
\begin{center}
\centerline{\includegraphics[width=2\columnwidth]{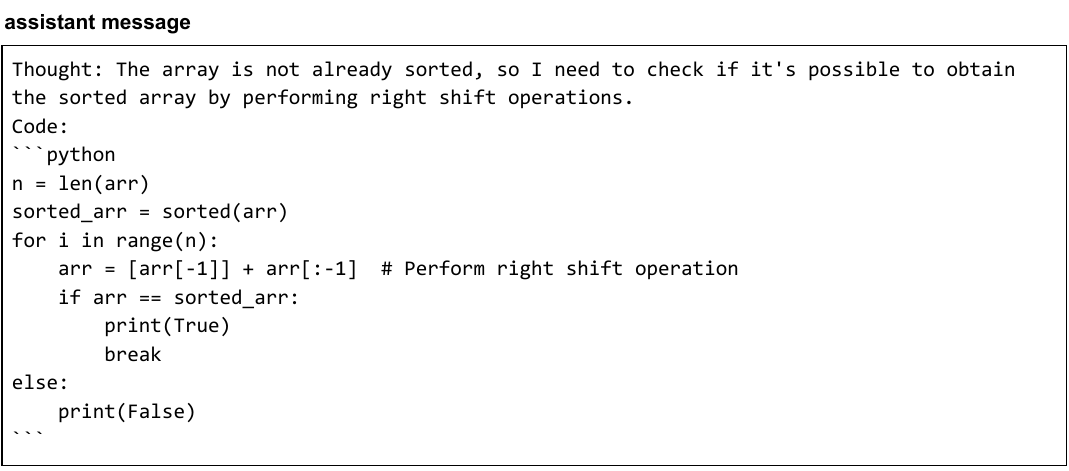}}
\caption{
Message 4 of Example 3.
}
\label{fig:example-3-4}
\end{center}
\vskip -0.2in
\end{figure*}

\begin{figure*}[ht]
\vskip 0.2in
\begin{center}
\centerline{\includegraphics[width=2\columnwidth]{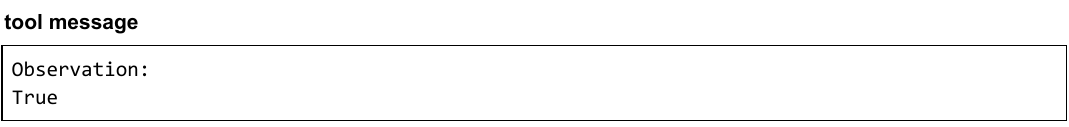}}
\caption{
Message 5 of Example 3.
}
\label{fig:example-3-5}
\end{center}
\vskip -0.2in
\end{figure*}

\begin{figure*}[ht]
\vskip 0.2in
\begin{center}
\centerline{\includegraphics[width=2\columnwidth]{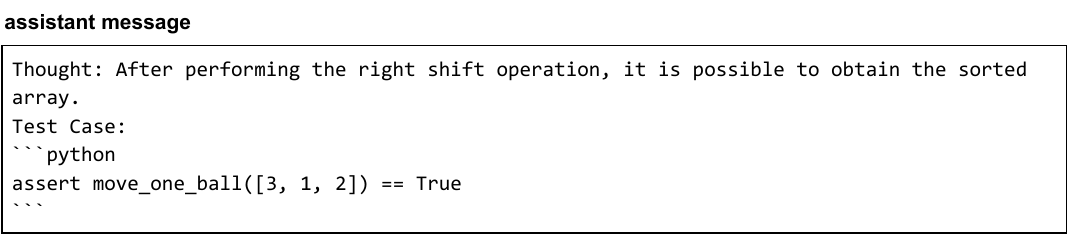}}
\caption{
Message 6 of Example 3.
}
\label{fig:example-3-6}
\end{center}
\vskip -0.2in
\end{figure*}

\begin{figure*}[ht]
\textbf{Example 4.}
\vskip 0.2in
\begin{center}
\centerline{\includegraphics[width=2\columnwidth]{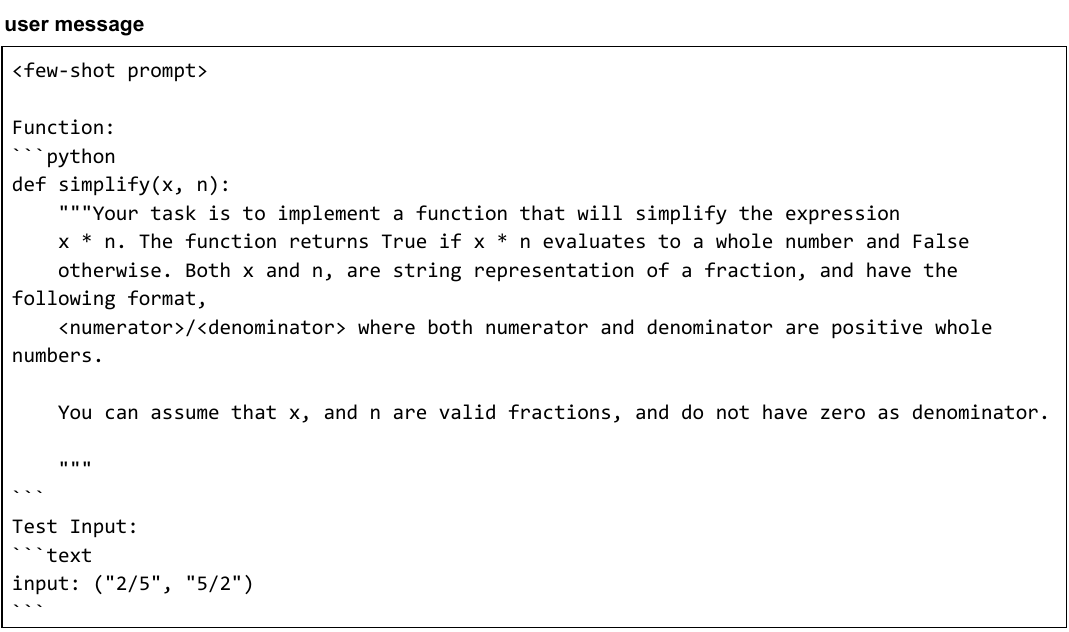}}
\caption{
Message 1 of Example 4.
}
\label{fig:example-4-1}
\end{center}
\vskip -0.2in
\end{figure*}

\begin{figure*}[ht]
\vskip 0.2in
\begin{center}
\centerline{\includegraphics[width=2\columnwidth]{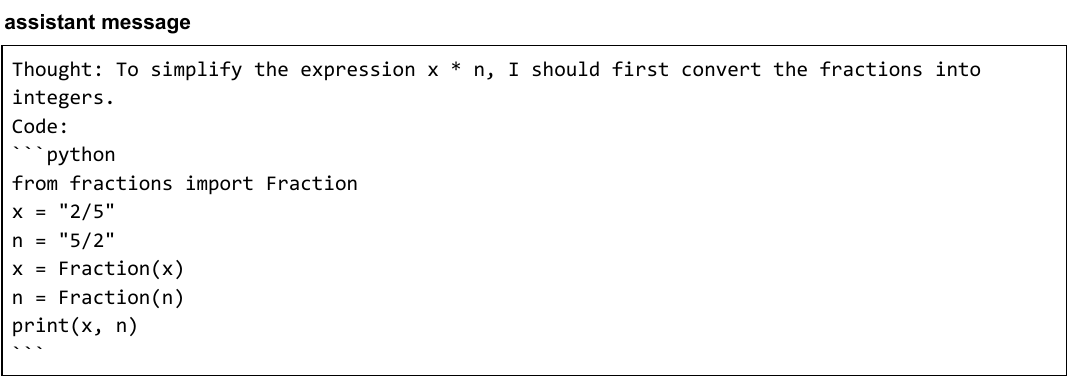}}
\caption{
Message 2 of Example 4.
}
\label{fig:example-4-2}
\end{center}
\vskip -0.2in
\end{figure*}

\begin{figure*}[ht]
\vskip 0.2in
\begin{center}
\centerline{\includegraphics[width=2\columnwidth]{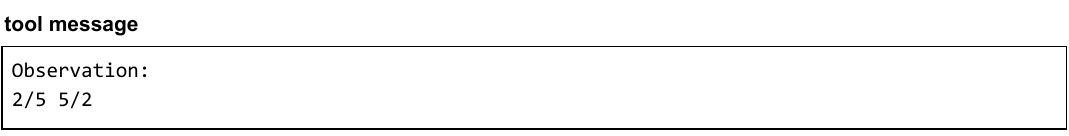}}
\caption{
Message 3 of Example 4.
}
\label{fig:example-4-3}
\end{center}
\vskip -0.2in
\end{figure*}

\begin{figure*}[ht]
\vskip 0.2in
\begin{center}
\centerline{\includegraphics[width=2\columnwidth]{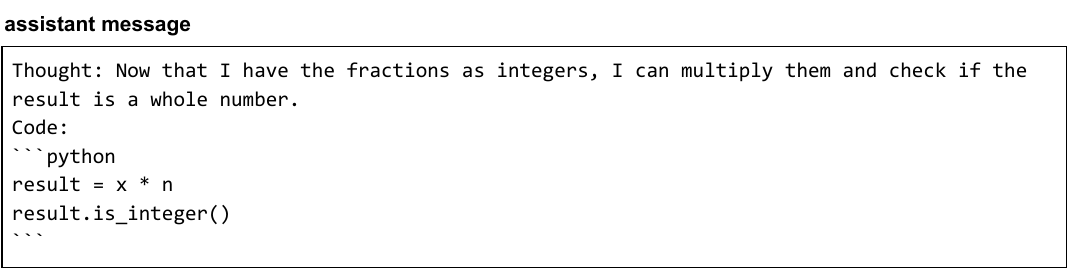}}
\caption{
Message 4 of Example 4.
}
\label{fig:example-4-4}
\end{center}
\vskip -0.2in
\end{figure*}

\begin{figure*}[ht]
\vskip 0.2in
\begin{center}
\centerline{\includegraphics[width=2\columnwidth]{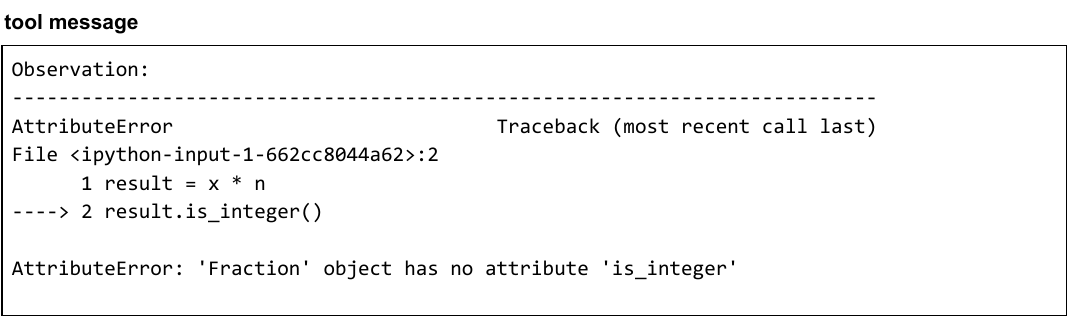}}
\caption{
Message 5 of Example 4.
}
\label{fig:example-4-5}
\end{center}
\vskip -0.2in
\end{figure*}

\begin{figure*}[ht]
\vskip 0.2in
\begin{center}
\centerline{\includegraphics[width=2\columnwidth]{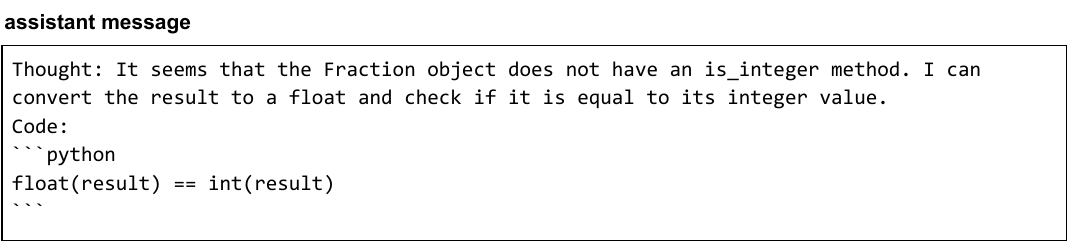}}
\caption{
Message 6 of Example 4.
}
\label{fig:example-4-6}
\end{center}
\vskip -0.2in
\end{figure*}

\begin{figure*}[ht]
\vskip 0.2in
\begin{center}
\centerline{\includegraphics[width=2\columnwidth]{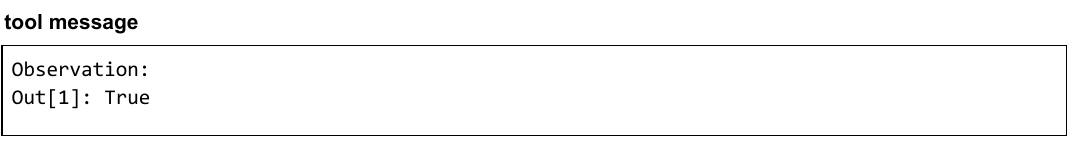}}
\caption{
Message 7 of Example 4.
}
\label{fig:example-4-7}
\end{center}
\vskip -0.2in
\end{figure*}

\begin{figure*}[ht]
\vskip 0.2in
\begin{center}
\centerline{\includegraphics[width=2\columnwidth]{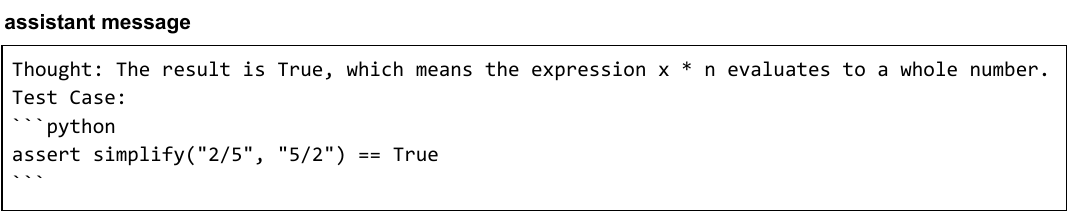}}
\caption{
Message 8 of Example 4.
}
\label{fig:example-4-8}
\end{center}
\vskip -0.2in
\end{figure*}

Figures~\ref{fig:example-1-1}--\ref{fig:example-4-8} provide four examples of
trajectories produced by Calculator agent, where the correct test outputs are 
obtained given the test inputs.

%These are test cases successfully generated by the input-output generator with a Python interpreter.

\end{document}